\documentclass[useAMS,usenatbib]{mnras}
\usepackage{graphicx,verbatim}
\usepackage[normalem]{ulem}
\usepackage{xcolor}
\usepackage[]{inputenc,amssymb}

\def \be{\begin{equation}}
\def \ee{\end{equation}}
\newcommand       \ba           {\begin{eqnarray}}
\newcommand       \ea           {\end{eqnarray}}
\def \bea{\begin{eqnarray}}
\def \eea{\end{eqnarray}}

\newcommand{\comments}[1]{}

\definecolor{webgreen}{rgb}{0,.5,0}
\definecolor{webbrown}{rgb}{.6,0,0}

\title[Spherical accretion on neutron stars and black holes]{Spherical accretion: the influence of inner boundary and quasi-periodic oscillations}   

\author[P. Dhang, P. Sharma, B. Mukhopadhyay]
{Prasun Dhang,$^{1}$\thanks{E-mail:prasundhang@gmail.com}
Prateek Sharma,$^{1}$
Banibrata Mukhopadhyay$^{1}$
\\
$^{1}$Department of Physics and Joint Astronomy Programme, Indian Institute of Science, Bangalore, INDIA 560012
}

\begin{document}
\maketitle

\label{firstpage}

\begin{abstract}
Bondi accretion assumes that there is a sink of mass at the center -- which in case of a black hole (BH) corresponds to the advection of matter 
across the event horizon. 
Other stars, such as a neutron star (NS), have surfaces and hence the infalling matter has to slow down at the 
surface. 
We study the initial value problem in which the matter distribution is uniform and at rest at $t=0$. 
We consider different inner boundary
conditions for BHs and NSs: outflow boundary condition (mimicking mass sink at the center) valid for BHs; and {\em reflective} and steady-shock 
(allowing gas to cross the inner boundary at subsonic speeds) boundary conditions for NSs. 
We also obtain a similarity solution for cold accretion on to BHs and NSs.
1-D simulations show the formation of an outward propagating and a standing shock in NSs for reflective and steady-shock boundary 
conditions, respectively. Entropy is the highest at the bottom of the subsonic region for reflective boundary conditions. 
In 2-D this profile is convectively unstable.
Using steady-shock inner boundary conditions, the flow is unstable to the standing accretion shock instability (SASI) in 2-D, which leads to
global shock oscillations and may be responsible for quasi-periodic oscillations (QPOs) seen in the lightcurves of accreting systems. 
For steady accretion in the quiescent state, spherical accretion rate on to a NS 
can be suppressed by orders of magnitude compared to that on to a BH.
\end{abstract}


\section{Introduction}           
\label{sect:intro}
Spherical accretion of adiabatic gas without angular momentum in steady state is the simplest model of accretion onto a central 
gravitating object, as proposed by \citet{Bondi1952}. The theory of Bondi accretion was studied numerically in a series of papers by 
Ruffert and his collaborators (\citealt{Ruffert1994a, Ruffert1994b, Ruffert1994c}). While spherical symmetry and adiabaticity are 
extreme idealizations -- angular momentum,  anisotropy due to large scale magnetic fields, entropy generation due to viscous/magnetic 
dissipation and conduction are almost always going to be important (\citealt{Shvartsman1971,Blandford1999,Sharma2008}) -- Bondi 
solution is the starting point for estimating the accretion rate in hot, non-radiative accretion flows on to BHs 
(\citealt{Baganoff2003,Loewenstein2001,Allen2006}). 

We recall that the Bondi solution assumes a uniform, static gas with a specified density and temperature at large distance from the 
accreting object. The Bondi radius ($\sim 2 GM/c_{s\infty}^2$; $c_{s\infty}$ is the sound speed in the ambient medium far away from 
the accreting object; $M$ is the mass of the central accretor) is a measure of the sphere of influence of the central gravitating object. 
The transonic solution, applicable for a 
black hole (BH) that allows mass to flow across the event horizon at the speed of light, becomes supersonic inside a critical radius 
($r_{c}$). Neutron stars (NSs), white dwarfs and normal stars, on the other hand, possess a surface and the accretion flow must 
necessarily attain zero or a very subsonic velocity at the stellar surface (strong cooling can, however, allow matter to fall freely on to the surface). 
Thus, the mass accretion rate and the gravitational energy released in the same ambient conditions may be vastly different 
for BHs and NSs.

Observations of BH and NS X-ray binaries (XRBs) show some dissimilarities. First, the BH XRBs are more luminous compared to 
NS XRBs in the high-soft/thermal state; this is expected as the maximal accretion rate (known as the Eddington rate) is proportional 
to the compact object mass (BHs are more massive than NSs). Second, the NS XRBs are observed to be more luminous compared 
to BH XRBs in the quiescent state (\citealt{Narayan1997,Garcia2001}; see, however, \citealt{Chen1998}). This difference between the 
BH and NS systems is often explained as follows. In quiescent state, owing to a low accretion rate, the accretion flow around the central 
compact object is optically thin and geometrically thick. It is in the hot, radiatively inefficient regime (e.g., \citealt{Das2013} and 
references therein). The gravitational power extracted via accretion is given by $\eta \dot{M} c^2$, where $\eta$ is the accretion 
efficiency and $\dot{M}$ is the accretion rate on to the compact object. For a BH accreting in the quiescent state the {\em radiative} 
output (e.g., in X-rays) is sub-dominant, and most of the accretion energy (stored as thermal/kinetic energy) is lost because of 
advection across the event horizon. However, advection is absent in a NS because of the surface, and all of the energy is 
thermalized and radiated ultimately (\citealt{Narayan1995,Narayan1997,Mukhopadhyay2002}).

The above argument implicitly assumes that the accretion rate on to the compact object ($\dot{M}$) is the same, irrespective of the 
central object. This assumption is likely to break down because of qualitatively different inner boundary conditions for BHs and NSs.
Studying BH and NS binaries with similar orbital periods, \citet{Menou1999} (see also \citealt{Asai1998}) showed that even if the 
mass transfer rates are similar (this is assumed for BHs and NSs with similar orbital periods), the accretion rate on to NSs is less 
compared to that on to BHs (this is inferred from the lower X-ray luminosity in NS XRBs compared to what is expected if all the accretion energy 
is radiated). They invoked the propeller effect (\citealt{ Illarionov1975}) due to a magnetosphere to suppress NS accretion relative to 
a BH in similar ambient conditions. A rotating magnetosphere flings away matter, preventing it from reaching the surface. However, 
not all quiescent NS binaries have strong enough magnetic fields for this to work (e.g., \citealt{DAngelo2015}). A floor in quiescent 
NS X-ray luminosity, which is larger than the most quiescent BHs, may also be explained if the X-ray emission is not due to current 
accretion, but say due to thermal radiation from a hot NS (e.g., \citealt{Cackett2010}), shock driven by an underlying pulsar 
(\citealt{Campana1998}), or due to coronal emission of the companion (e.g., \citealt{Bildsten2000}).

The most prominent difference between NSs and BHs is that, unlike the latter, the former have a hard surface. We want to isolate the effect 
of a hard surface in NS accretion. Therefore, we consider a set-up without magnetic fields and angular momentum. While this is very 
idealized, it illustrates the essential (and potentially large) differences between BH and NS accretion in the quiescent state. We show that,
in similar ambient conditions, the mass accretion rate on to NSs compared to BHs in the radiatively inefficient regime can be orders of 
magnitude lower. Thus, one cannot simply use the presence of a hard surface in NSs to explain their larger radiative output in the quiescent 
state.

An outcome of our idealized
simulations is that for some inner boundary conditions we obtain large amplitude coherent oscillations of a standing shock. We identify
these oscillations as a result of the standing accretion shock instability or SASI, widely seen in multi-dimensional supernova simulations 
(e.g., see \citealt{Hanke2012} and references therein). We propose that the coherent oscillations seen in our simulations may be responsible 
for some of the quasi-periodic oscillations (QPOs) observed in the lightcurves of XRBs. The observed QPOs have frequencies ranging from mHz (low-frequency 
QPOs) to kHz (kHz QPOs). Some models describe the observed variability based on the orbital and epicyclic motions, while some are based on resonance 
between disk rotation and NS spin. Some models identify QPOs as global disk oscillation modes. 
Although several models have been proposed over the years, the mechanism(s) responsible for QPOs is still unknown 
(for a review, see \citealt{vanderklis2004}). 

The paper is organized as follows. In section \ref{sect:set up} we describe our physical set-up. In section \ref{sect:similarity} we describe 
the similarity solutions for cold spherical accretion with outflow and reflective inner boundary conditions. In section \ref{sect:simulation} 
we describe the results from our numerical simulations. In section \ref{sect:discussion} we discuss the astrophysical implications of our 
results and summarize in section \ref{sect:conclusions}. 

\section{Physical set-up}
\label{sect:set up}
We set up an initial value problem in which a gravitating compact object is embedded in an initially static, spherically-symmetric medium. 
In section \ref{sect:gen_sim} we consider similarity solutions for compact objects embedded in static atmospheres with power-law density.
We take the central gravitating objects to be either a black hole (BH) or a neutron star (NS). We solve the Euler equations  
to study the problem of spherical accretion on to NSs and BHs. The only difference between NS and BH accretion in our formalism is that we 
impose reflective boundary condition (or a steady-shock boundary condition; see section \ref{sect:ini_bdry}) for the flow at the inner boundary 
in former and outflow boundary condition in latter.  The Euler equations in presence of gravity of a compact object are given by
 \ba 
 \label{eq:mass}
&& \frac{\partial \rho}{\partial t} + \nabla .(\rho \textbf{v})= 0, \\
\label{eq:momentum}
&& \frac{\partial }{\partial t} \left ( \rho \textbf{v} \right )+ \nabla. \left( \rho \textbf{v}\textbf{v} + p\textbf{I} \right) = - \rho \nabla \Phi, \\
\nonumber
&& \frac{\partial} {\partial t} \left( \frac{\rho \textbf{v}^2}{2} + e +\rho \Phi \right ) + \nabla . \left[ \textbf{v} \left ( \frac{\rho \textbf{v}^2}{2} 
+ e+p +\rho \Phi \right ) \right] = 0, \\
\label{eq:energy}
&&
\ea
where, $\rho$, $\textbf{v},~p~(=[\gamma-1]e),~e$ are the mass density, velocity, pressure, and internal energy density, respectively, 
and $\gamma$ (chosen to be 1.4, unless mentioned otherwise) is the adiabatic index. 
The Newtonian potential of the central accretor is $\Phi = -{GM}/{r}$, where $G$ is Newton's gravitational constant, $M$ is the mass of the central compact object, 
and $r$ is the distance from the center. We use Newtonian potential because the similarity solution that we obtain in section \ref{sect:similarity} is strictly applicable
only for a scale-free potential. Newtonian approximation should not change our results qualitatively; the differences between BH and NS accretion are mostly due 
to the presence of a hard surface in latter (which we mimic by using a different inner boundary condition from BHs). We have verified that the results from 
numerical simulations using a pseudo-Newtonian potential mimicking general relativistic effects (\citealt{Paczynsky1980}) are qualitatively similar to the 
Newtonian simulations. This is essentially because the inner boundary affects the flow far away from the event horizon, where GR effects are
subdominant. 

\section{Similarity Solution}
\label{sect:similarity}

In the limit that the external medium (loosely referred to as interstellar medium or ISM) is cold ($p_{\rm ISM}=0$), the only length scale that can 
be constructed from the parameters of the problem is
  \begin{equation}
  \label{eq:rs}
    r_{s}(t) = \beta (\gamma) (GMt^2)^{1/3},
  \end{equation}
where $\beta(\gamma)$ is a function of the adiabatic index ($\beta \approx 0.21$ for $\gamma=1.4$ and it increases with an increasing $\gamma$). 
This scale radius applies to both NSs and BHs but the solutions in these two cases differ because the 
inner boundary conditions in the two cases are different. While the accretion flow comes to rest at the center ($v=0$ at $r=0$) for NSs, the matter is freely-falling 
($v=\sqrt{GM/r}$ as $r \to 0$) on to BHs. Since matter falls freely in a cold medium and the dynamical time is shortest at the center, the supersonic 
matter coming to rest at NS surface ($r=0$) launches an outward propagating shock.  For a BH, matter is removed supersonically from the inner boundary and
there is no shock. Eq. \ref{eq:rs} can be taken as the shock radius as a function of time for a NS, and as a scale radius for a BH. In both cases, the solutions
(density, pressure, velocity profiles) as a function of distance from the center at different times can be scaled to $r_s(t)$. The solutions as a function of the 
scaled radius (defined in Eq. \ref{eq:sim_var}) lie on top of each other when scaled appropriately (Eq. \ref{eq:scaling} or \ref{eq:scaling2}).

The velocity with which the scale radius (equal to the shock velocity for a NS) moves out is 
\begin{equation}
\label{eq:vs}
v_s\equiv \frac{dr_s}{dt} = \frac{2r_s}{3t}= \frac{2}{3} \beta \left (\frac{GM}{t} \right)^{1/3}.
\end{equation}
We can convert the partial differential equations (PDEs; Eqs. (\ref{eq:mass})-(\ref{eq:energy}) into ordinary differential equations (ODEs) if we scale the length scales with $r_s$ and velocities with $v_s$. This is the essence of the similarity method, the well-known application of which is the Sedov-Taylor solution for a point explosion (\citealt{Sedov1946,Taylor1950}). We introduce a similarity variable 
  \begin{equation}
    \label{eq:sim_var}
    \xi = \frac{r}{r_s(t)},
  \end{equation}
 which captures both the spatial and temporal evolution. We discuss the similarity solutions for two different initially static density profiles: i) a uniform density, 
 $\rho = \rho_0$; and ii) a power law density, $\rho = Dr^{-\alpha}$ ($D,~\alpha$ are parameters).
 
\subsection{Uniform density}
\label{sect:uniform}
The scaled density, velocity and pressure are as follows:
  \begin{equation}
    \label{eq:scaling}
    \tilde{\rho} = \frac{\rho}{\rho_{0}} ,~\tilde{v} = \frac{v}{v_s},~\tilde{p} = \frac{p}{\rho_{0}v_s^2},
   \end{equation}
where $\rho_0$ is the initial density of the uniform ambient medium and $v_s$ is the velocity corresponding to the scaling radius. Plugging these 
scalings in 1-D spherically symmetric form of  the mass and momentum equations (Eqs. \ref{eq:mass} and \ref{eq:momentum}), we obtain
  \ba
    \label{eq:mass_sim}
     &&(-\xi + \tilde{v})\frac{d\tilde{\rho}}{d\xi} + \tilde{\rho}\frac{d\tilde{v}}{d\xi} + \frac{2}{\xi}\tilde{\rho}\tilde{v} = 0,\\
    \label{eq:mom_sim}
    &&(-\xi + \tilde{v})\frac{d\tilde{v}}{d\xi} + \frac{1}{\tilde{\rho}}\frac{d\tilde{p}}{d\xi} = \frac{\tilde{v}}{2} - \frac{9}{4\beta^3}\frac{1}{\xi^2}.
  \ea
Since the entropy of a fluid element is conserved, except when it crosses the shock (which forms in case of a NS), the alternative form of the energy equation (Eq. \ref{eq:energy}) is
  \begin{equation}
    \label{eq:entropy}
    \left(\frac{\partial}{\partial t} + v\frac{\partial}{\partial r}\right)\frac{p}{\rho ^{\gamma}} = 0,
      \end{equation}
which, when scaled, becomes
  \begin{equation}
    \label{eq:ent_sim}
     (-\xi+v) \frac{d}{d\xi}\bigg(\frac{\tilde{p}}{\tilde{\rho}^\gamma}\bigg) - \frac{\tilde{p}}{\tilde{\rho}^\gamma}=0.
  \end{equation}
Note also that this equation is valid everywhere except at the shock. Also note that Eqs. \ref{eq:mass_sim}, \ref{eq:mom_sim} and \ref{eq:ent_sim} 
are ordinary differential equations in one independent variable ($\xi$), much simpler to solve than the original partial differential equations. 
The spatial and temporal dependence of physical quantities is obtained by the scaling relations in Eq. (\ref{eq:scaling}).

        \begin{figure}
	\includegraphics[scale=0.4]{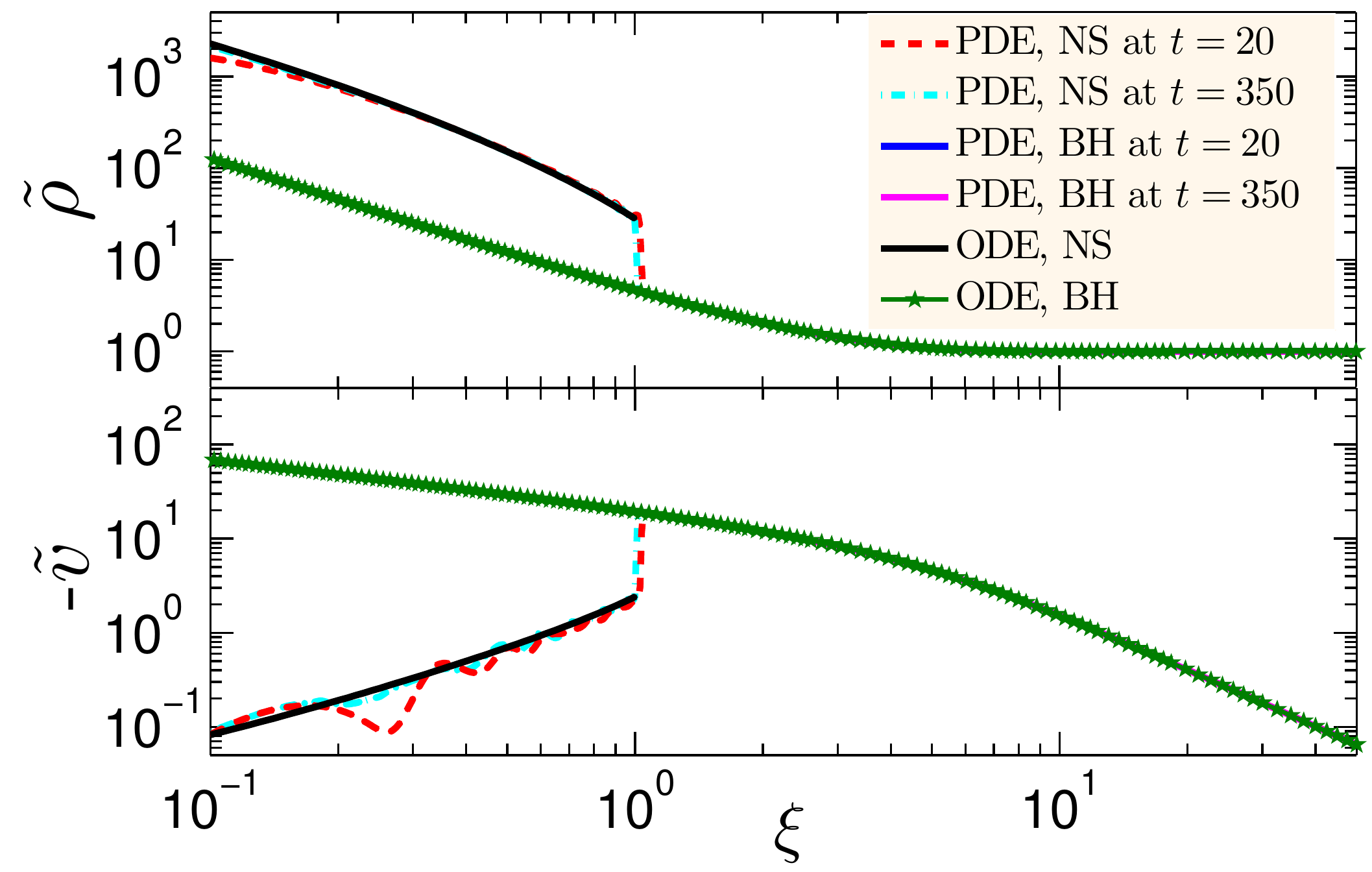}
	\caption{Scaled density ($\tilde{\rho}$; top panel) and velocity ($\tilde{v}$; bottom panel) profiles for the similarity (ODE) and  PDE 
	solutions with a BH and a NS. While all scaled BH solutions lie on top of each other, the NS PDE solutions show slight deviations from the ODEs 
	because of small numerical errors. Note that the velocity is infalling everywhere. Time $t$ is in the units of $r_g/c$.}  
		\label{fig:compare}
         \end{figure}
For a cold interstellar medium (ISM), far away from the accreting object ($\xi \to \infty$), $\tilde{p}=0$ and $\tilde{\rho}$ is unity (scaling the density to the ambient value). Eqs. (\ref{eq:mass_sim}) and (\ref{eq:mom_sim}) as $\xi \to \infty$ give $\tilde{v}_{\infty} = -3/(2 \beta^3 \xi^2)$. These are chosen as the outer boundary conditions for the ODEs (Eqs. \ref{eq:mass_sim}, \ref{eq:mom_sim}, and \ref{eq:ent_sim}). For BHs, the pressure remains zero throughout; i.e., $\tilde{p}=0$. Therefore, for BHs we simply integrate Eqs.  \ref{eq:mass_sim} and \ref{eq:mom_sim} inwards.
In contrast,  for NSs a shock forms, within which pressure roughly balances gravity. The inner boundary condition for NSs is chosen to be $v\to 0$ as $\xi \to 0$; i.e.,
flow comes to rest at the surface. Moreover, Rankine-Hugoniot shock jump conditions are applied across the shock at $\xi=1$. For a strong shock, 
the post-shock quantities (denoted by subscript 2) are related to the pre-shock quantities (denoted by subscript 1) as 
(note that the quantities are measured in a frame in which the shock is not at rest)
   \begin{equation}
   \label{eq:RH}
        \tilde{v}_2 = \frac{\gamma-1}{\gamma+1}\tilde{v}_1 + \frac{2}{\gamma+1}, ~\tilde{\rho}_2=\frac{\gamma+1}{\gamma-1}\tilde{\rho}_1,
         ~\tilde{p}_2=\frac{2 \tilde{\rho}_1[1-\tilde{v}_1]^2}{\gamma+1}.
    \end{equation}
We compare the results that we get from similarity solution (ODE) with the PDE solutions discussed later. Fig. \ref{fig:compare} shows a comparison between the ODE and PDE solutions for density and velocity at different times. We see that density profiles at different times coincide, so do the velocity profiles at different times. This is a signature of similarity solution which we have stated before also. While the match between the ODE and PDE solutions is perfect for BH, there are small discrepancies for NS solutions because of small errors at the inner boundary. As expected, the NS and BH solutions match outside the shock.

\subsection{Power-law density}
\label{sect:gen_sim}
In this section we consider a power-law initial density profile for the cold ISM, $\rho_{\rm ini} = D r^{-\alpha}$. We can obtain the similarity solution following 
a procedure similar to section \ref{sect:uniform}. The scale length, shock velocity and similarity variable are still given by Eqs. \ref{eq:rs}-\ref{eq:sim_var}. 
The scaled density, velocity, and pressure in this case are
    \begin{equation}
        \label{eq:scaling2}
        \tilde{\rho} = \frac{\rho}{Dr_s^{-\alpha}} ,~\tilde{v} = \frac{v}{v_s},~\tilde{p} = \frac{p}{Dr_s^{-\alpha}v_s^2}.
    \end{equation}
    
          \begin{figure}
	\includegraphics[scale=0.4]{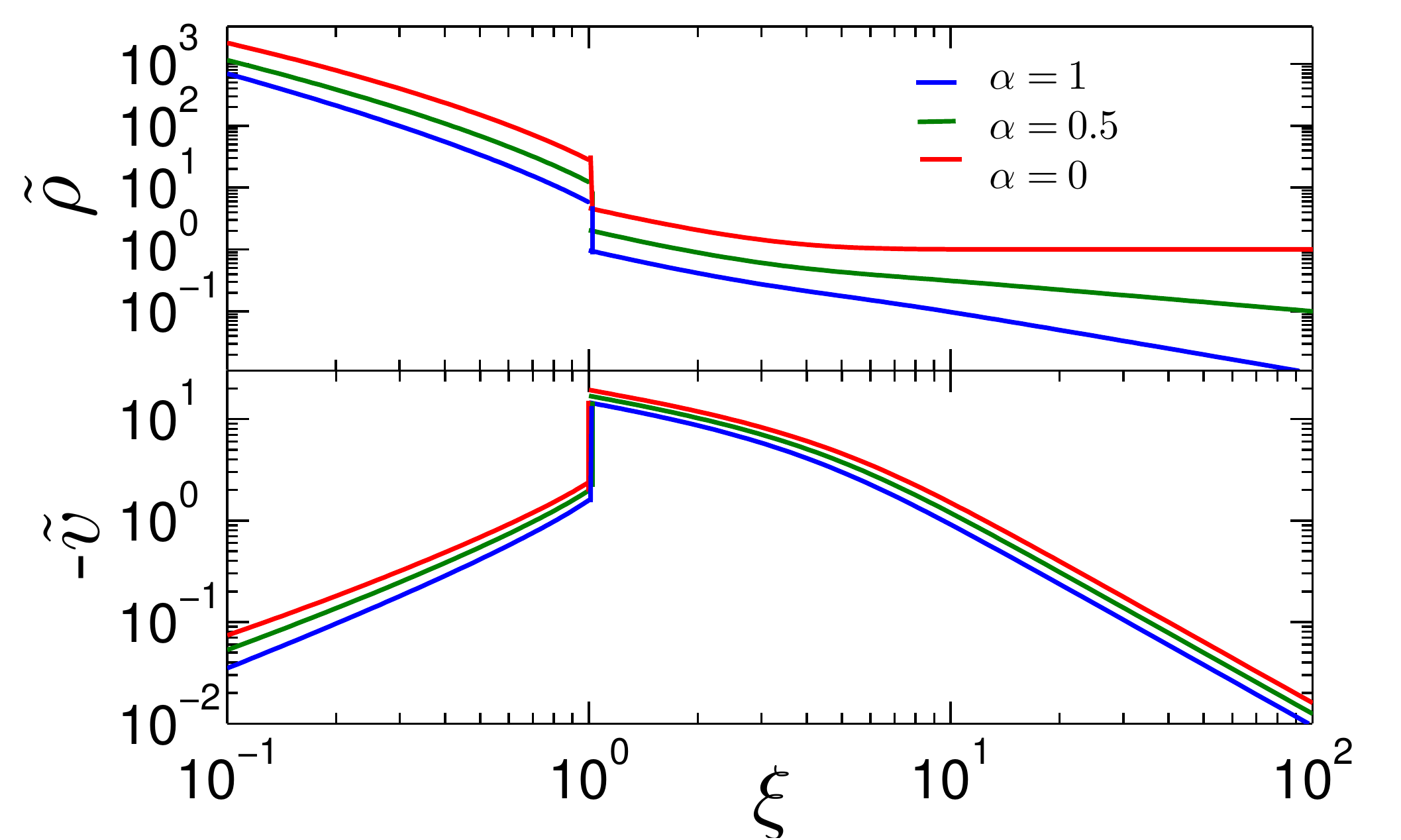}
	\caption{Scaled density ($\tilde{\rho}$; top panel) and velocity ($\tilde{v}$; bottom panel) profiles for NS solutions with 
	different initial density power-law profiles ($\alpha$s; at $t=0$ $\rho \propto r^{-\alpha}$).}  
	\label{fig:alpha}
          \end{figure}
Using these scaled variables and considering spherical symmetry, the 1-D mass, momentum, and entropy equations (Eqs. \ref{eq:mass}, \ref{eq:momentum} and \ref{eq:entropy}) take the form
    \ba
        \label{eq:mass_sim2}
        &&(-\xi + \tilde{v})\frac{d\tilde{\rho}}{d\xi} + \tilde{\rho}\frac{d\tilde{v}}{d\xi} + \frac{2}{\xi}\tilde{\rho}\tilde{v} - \alpha \tilde{\rho} = 0,\\
        \label{eq:mom_sim2}
        &&(-\xi + \tilde{v})\frac{d\tilde{v}}{d\xi} + \frac{1}{\tilde{\rho}}\frac{d\tilde{p}}{d\xi} = \frac{\tilde{v}}{2} - \frac{9}{4\beta^3}\frac{1}{\xi^2},\\
        \label{eq:ent_sim2}
        &&(-\xi+\tilde{v}) \frac{d}{d\xi}\bigg(\frac{\tilde{p}}{\tilde{\rho}^\gamma}\bigg) + [(\gamma-1)\alpha -1] \frac{\tilde{p}}{\tilde{\rho}^\gamma}=0.
     \ea
Using a method similar to section \ref{sect:uniform}, we solve Eqs. (\ref{eq:mass_sim2}) - (\ref{eq:ent_sim2}) to obtain the similarity solution for different density 
power laws ($\alpha$s). The boundary conditions as $\xi \to \infty$ are  $\tilde{\rho}_\infty = \xi^{-\alpha}$ and $\tilde{v}_\infty = -3/(2\beta^3 \xi^2)$.
Similarly, for NSs, $\tilde{v} \to 0$ as $\xi \to 0$ is the additional boundary condition. Moreover, for NSs the shock jump conditions in Eq. \ref{eq:RH} apply
across the shock at $\xi=1$. 

Fig. \ref{fig:alpha} shows the scaled density and velocity profiles. We also find that these results are in good agreement with the results obtained 
from	the numerical simulations of PDEs. Note that $\beta$ (see Eq. \ref{eq:rs}) for a power-law density profile depends on $\alpha$, the power-law 
slope, and therefore the velocity at large radii $(\tilde{v}_\infty = -3/[2\beta^3 \xi^2])$ differ for different $\alpha$s.

We note that  \citet{sakashita1974a} and \citet{sakashita1974b} have also obtained similarity solutions for NSs with shocks. Before these, 
\citet{Bisnovatyi-Kogan1972} obtained similarity solutions with shocks in a uniform gravitational field. However, there are
several differences from our approach. These early papers do not consider the solution outside the shock; specifically, the large $\xi$ regime 
in which $\tilde{v}_\infty = -3/(2 \beta^3 \xi^2)$ is missing. Also, rather than solving ODEs at all radii, these analytic works
assume an asymptotic power-series form of profiles for $\xi < 1$. Not only do we obtain the correct solutions for the profiles at all $\xi$, we 
also show that similar solutions are applicable for BHs, in which shocks are absent. We have also confirmed our 
solutions with the numerical solutions of PDEs discussed in section \ref{sect:simulation}.

\section{Time-dependent numerical simulations}
\label{sect:simulation}

In this section we describe the time dependent numerical simulations of Euler equations (Eqs. \ref{eq:mass}-\ref{eq:energy}) treated as PDEs. We use the 
 {\tt PLUTO} hydrodynamic code in 1-D and 2-D (\citealt{Mignone2007}). The problem set-up is the same as described in section \ref{sect:set up}, 
 namely a static uniform medium at $t=0$.  We consider two kinds of ambient conditions: (i) a cold ISM to check the validity of our similarity solution in section \ref{sect:similarity}; (ii) a warm ISM, 
 in which steady state solutions for BHs and NSs are attained. The steady transonic solution for BHs is the well-known Bondi solution (\citealt{Frank2002}).

\subsection{Initial and boundary conditions}
\label{sect:ini_bdry}
The initial pressure is $p_0 =\rho_0 c_{s\infty}^2/\gamma$, where $c_{s\infty}=(\gamma p_0/\rho_0)^{1/2}$ is the  sound speed in the ambient ISM. 
As mentioned earlier, we consider initial conditions: (i) a cold ISM (with $c_{s\infty}=0$) for comparing with the similarity solution; and (ii) a warm ISM (with $c_{s\infty}^2=0.002 \gamma c^2$, where $c$ is the speed of light in vacuum) with the sonic radius $r_{c}=(5-3\gamma)GM/(4c_{s\infty}^2)$ (the radius at which the 
radial inflow velocity equals the local sound speed) and the Bondi radius 
$r_{B} = 8 r_c/(5-3\gamma)=2GM/c_{s\infty}^2 \approx 714 GM/c^2$ (for our choice of parameters) well inside the simulation domain, for which the 
PDE solution reaches a steady state.

\subsubsection{1-D}
\label{sect:1-D_boundary}
The 1-D runs are carried out till $10^6 r_g/c$, time by which a steady solution is attained by our warm ISM simulations.
In all cases computational domain extends from an inner boundary $r_{\rm in}=6r_g$ (corresponding to the innermost stable circular orbit of a Schwarzschild 
black hole and roughly the size of a neutron star) to an outer boundary $r_{\rm out}=10^4r_{g}$, where $r_{g}=GM/c^2$ is the gravitational radius. A logarithmic grid with $N_r=1024$ grid points is used. 
For both BHs and NSs, the outer boundary conditions are the same; we fix the pressure and density to their initial values $p_0$ and $\rho_0$, and set the velocity to 
zero. For BHs we use the outflow boundary condition at the inner boundary, such that mass is allowed to be advected into the BH but cannot 
come out of it. 

For most NS runs reflective boundary condition is applied at the inner radius, such that matter comes to rest at the stellar surface.
Another inner boundary condition that we use for some NS runs is what we call the `steady-shock' boundary condition (a similar boundary condition is used by \citealt{Blondin2003}, who call it a leaky boundary condition), in which the velocity of
the gas crossing the inner boundary ($v_{\rm in}$ in the ghost zones) is fixed to a given subsonic value (chosen to be $0.05$c). The values of pressure and 
density in the ghost zones are copied from the innermost zone of the computation domain. The importance of the inner boundary condition for NSs is discussed 
in section \ref{sect:discussion}.

\subsubsection{2-D}
We also carry out 2-D axisymmetric simulations in spherical $(r,\theta,\phi)$ coordinates for a warm ambient medium . The 2-D simulations are 
necessary because multi-dimensional effects such as convection and standing shock instability can qualitatively change the accretion flow.
The computational domain extends
 from an inner radius $r_{\rm in} = 6r_g$ to an outer radius $r_{\rm out} = 10^4r_g$. We use a logarithmic grid along radial direction with the number of grid 
 points $N_r = 512$. Along $\theta$ ($0 \leq \theta \leq \pi$) a uniform grid with $N_\theta = 128$ is used. 

Along the radial direction we use the same outer boundary conditions as in 1-D simulations (see section \ref{sect:1-D_boundary}). 
Additionally, we set all velocity components to zero at the outer boundary. 
At the inner boundary we fix the tangential velocity ($v_{\theta}$) to zero for the  steady-shock boundary condition; similar results are obtained if $v_\theta$ is copied in
the ghost zones instead of being set to zero. 
The tangential velocity is copied
in the ghost zones at the inner radial boundary for both reflective (applicable to NSs) and outflow (applicable to BHs) boundary conditions.
In all cases, axisymmetric boundary conditions are used for both BHs and NSs at $\theta=0$ and $\theta=\pi$.\\

\subsection{Simulation results}
\subsubsection{1-D}
\label{sect:1d_result}
\begin{figure}
	\includegraphics[scale=0.4]{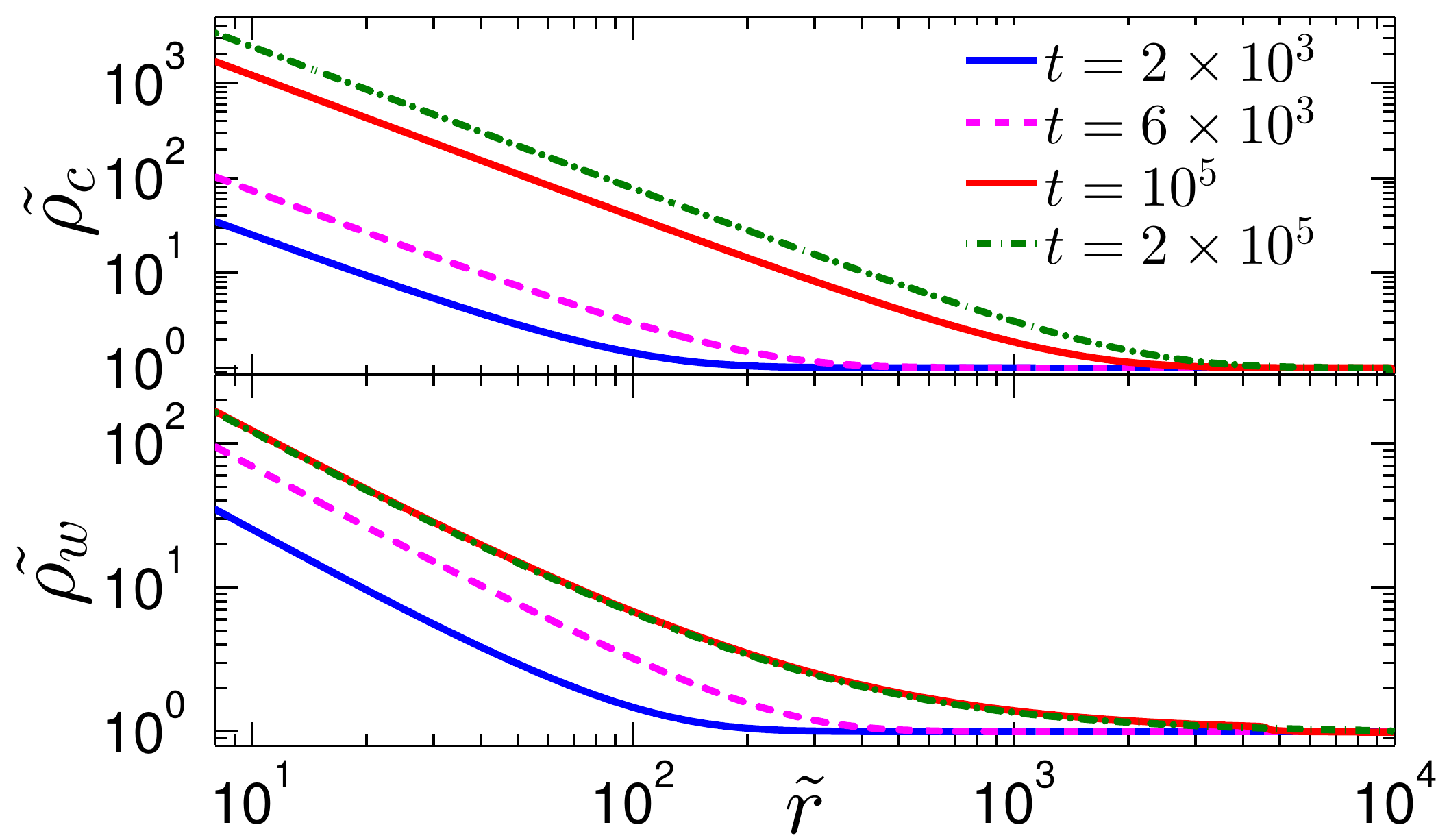}
	\caption{Density profiles at different times for cold ($\tilde{\rho}_{c}=\rho_c/\rho_0$; top panel) and warm ($\tilde{\rho}_{w}=\rho_{w}/\rho_0$; 
	bottom panel) ISM in 1-D BH simulations. 
	Time $t$ is in units of $r_g/c$.}
	\label{fig:den_BH_hot_cold}
\end{figure}

\begin{figure}
	\includegraphics[scale=0.4]{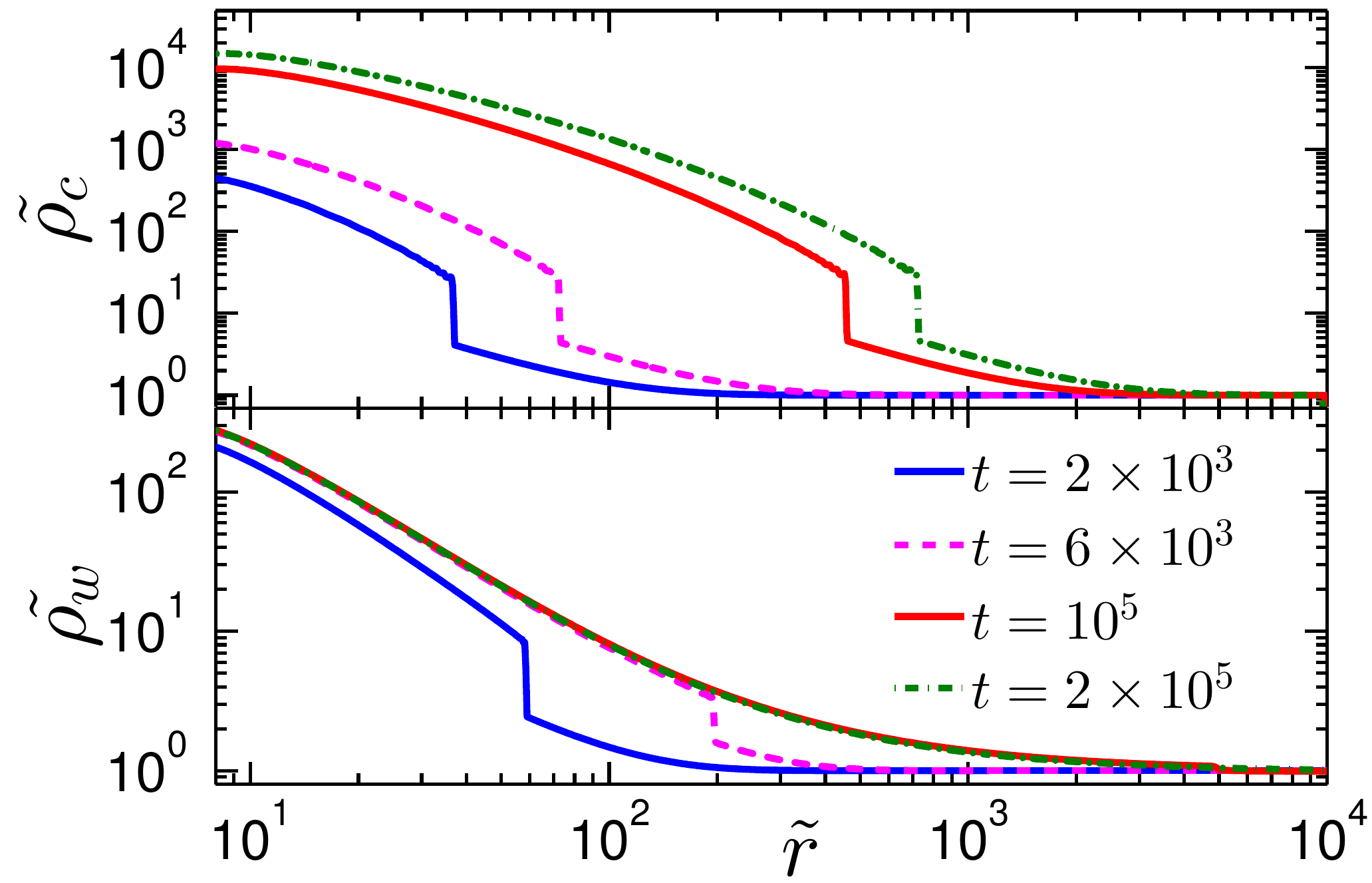}
	\caption{Density profiles at different times for cold ($\tilde{\rho}_{c}=\rho_c/\rho_0$; top panel) and warm ($\tilde{\rho}_{w}=\rho_{w}/\rho_0$; 
	bottom panel) ISM in 1-D NS simulations with reflective inner boundary condition.  	Time $t$ is in units of $r_g/c$.}
	\label{fig:den_NS_hot_cold}
\end{figure}

Fig. \ref{fig:den_BH_hot_cold} shows the density profiles for BH simulations at different times for an initially cold (top panel) and warm (bottom panel) ISM. 
The density increases with time for a cold ISM, never reaching a steady state; pressure, being zero, is never able to balance the gravitational 
pull and the density front keeps propagating out in accordance with Eq. \ref{eq:rs}. 
For a warm ISM, on the other hand, the density profile attains a steady state roughly after the density front reaches the Bondi radius (i.e., at $r_B/c_{s\infty} \approx 1.3 \times 10^4$). Velocity takes far longer time ($\sim r/v$) to equilibrate at large 
radii because the flow is extremely subsonic ($v \propto 1/r^2$) beyond the Bondi radius. In steady state, the warm ISM settles down to the classic Bondi 
solution.

Fig. \ref{fig:den_NS_hot_cold} shows the scaled density profiles for accretion on to a NS immersed in a cold (top panel) and a warm (bottom panel) ISM; 
 reflective boundary condition is used at the inner radius. For a NS with reflective boundary condition, an outward-propagating shock is launched. 
For a cold ISM, the shock is always strong as the shock Mach number is infinite, and the solution never reaches a steady state (like in the case of a BH). 
In contrast, for a warm ISM the shock weakens as its velocity becomes comparable to the upstream sound speed. Eventually the flow attains a steady state 
and there is no shock; this state is essentially a hydrostatic atmosphere.

\begin{figure}
	\includegraphics[scale=0.4]{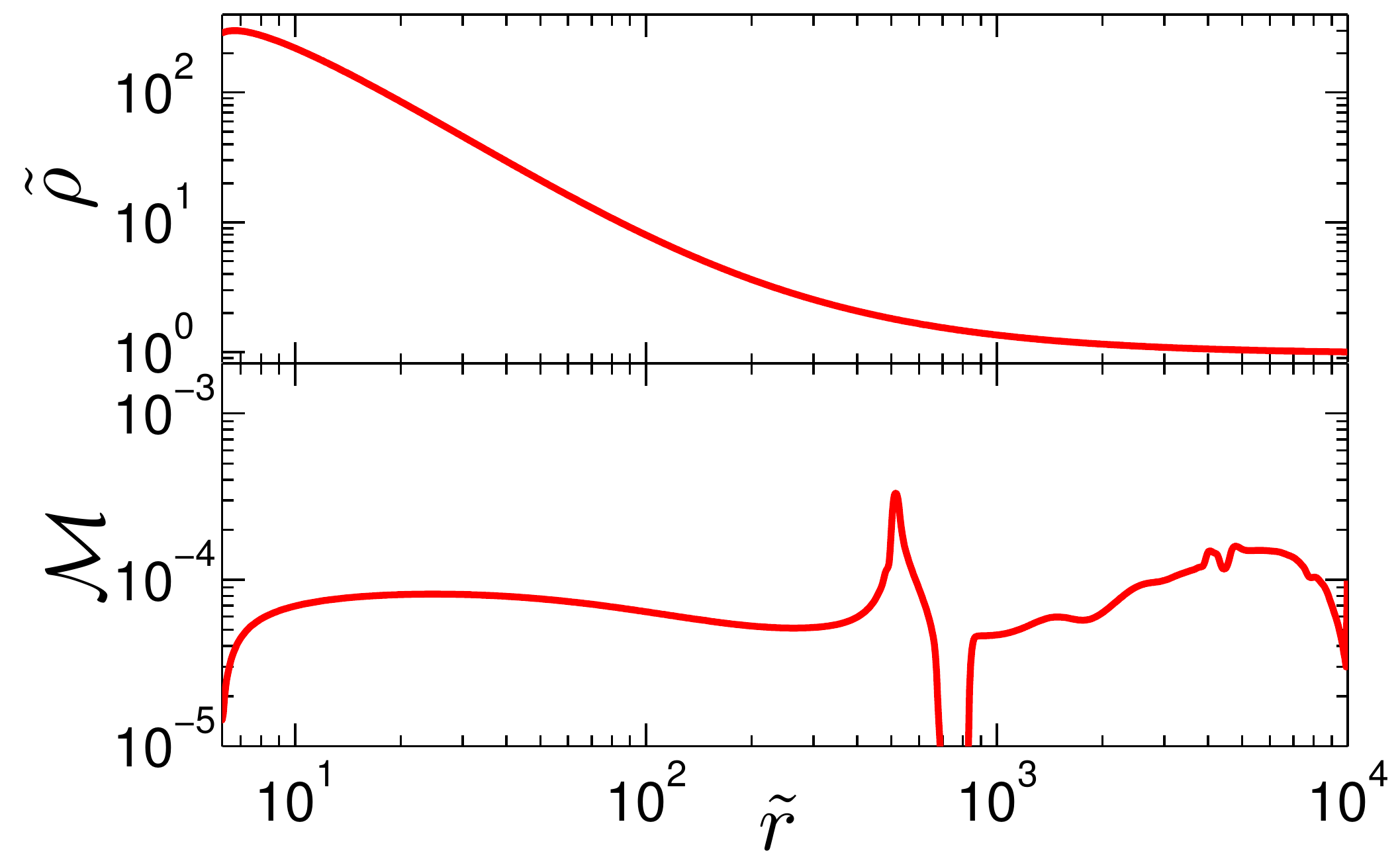}
	\caption{Density ($\tilde{\rho} = \rho/\rho_0$; top panel; same as the late-time profiles in the bottom panel of Fig. \ref{fig:den_NS_hot_cold}) and Mach 
	number (${\cal M} \equiv - v/c_{s}$, where $c_s\equiv [\gamma p/\rho]^{1/2}$ is the local sound speed; bottom panel) profiles in steady state for the warm 
	1-D NS run with reflective inner boundary condition.} 
	\label{fig:steady}
\end{figure}

Fig. \ref{fig:steady} shows the scaled density and Mach number (${\cal M} \equiv -v/c_s$; $c_s=[\gamma p/\rho]^{1/2}$ is the local sound speed) profiles in steady
state for the warm NS run with reflective inner boundary condition. The Mach numbers and velocities are tiny $\sim 10^{-4}$ throughout the box;  
at some radii the velocity is positive! Strictly speaking, in steady state, the velocity throughout the computational domain should be zero because the 
mass flux at the inner boundary is set to zero. Small velocities arise because of small numerical errors at the inner boundary (these errors decrease with increasing
resolution and at late times).

\begin{figure}
	\includegraphics[scale=0.39]{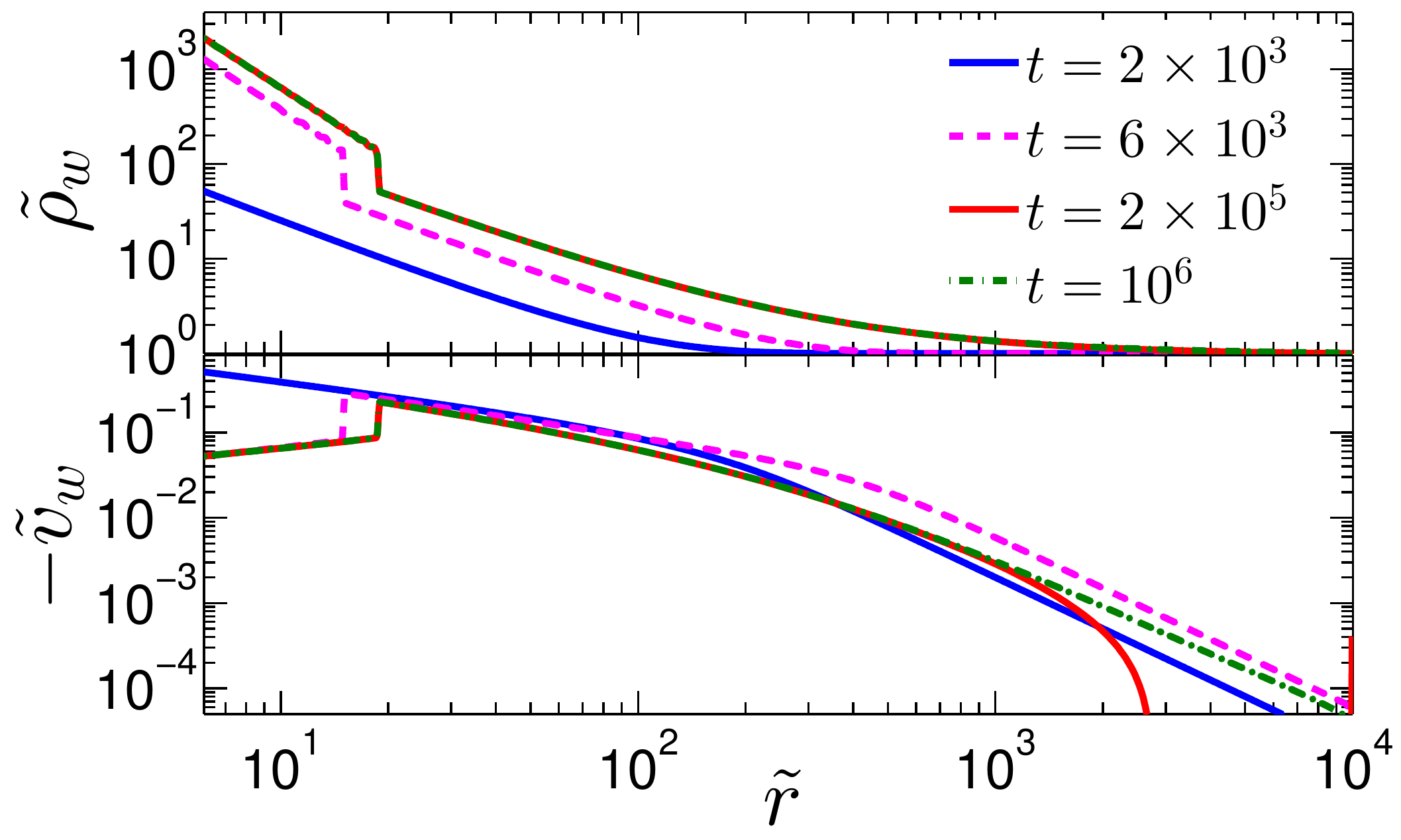}
	\caption{Density ($\tilde{\rho}_{w}=\rho_w/\rho_0$; top panel) and velocity profiles ($\tilde{v}_w = v_w/c$; bottom panel) at different times for 1-D 
	accretion on to  a NS with steady-shock boundary condition.} 
	\label{fig:den_steady}
\end{figure}

Fig. \ref{fig:den_steady} shows the scaled density and velocity profiles for accretion on to a NS in a warm medium with the steady-shock inner 
boundary condition (see section \ref{sect:1-D_boundary}). In this case too, an outward-propagating shock is launched initially, but it does not vanish in 
steady state, unlike what happens with the inner reflective boundary condition. Instead,  the shock halts and becomes a steady non-propagating/standing 
shock. In steady state the density and velocity profiles outside the shock are identical to the warm BH solutions (see bottom panel of Fig. 
\ref{fig:den_BH_hot_cold}), but within the shock the density is much 
higher and the flow is subsonic. Fig. \ref{fig:den_steady} shows that the velocity profiles at large radii indeed take much longer to attain a steady state as 
compared to the density profiles.

\begin{figure}
	\includegraphics[scale=0.42]{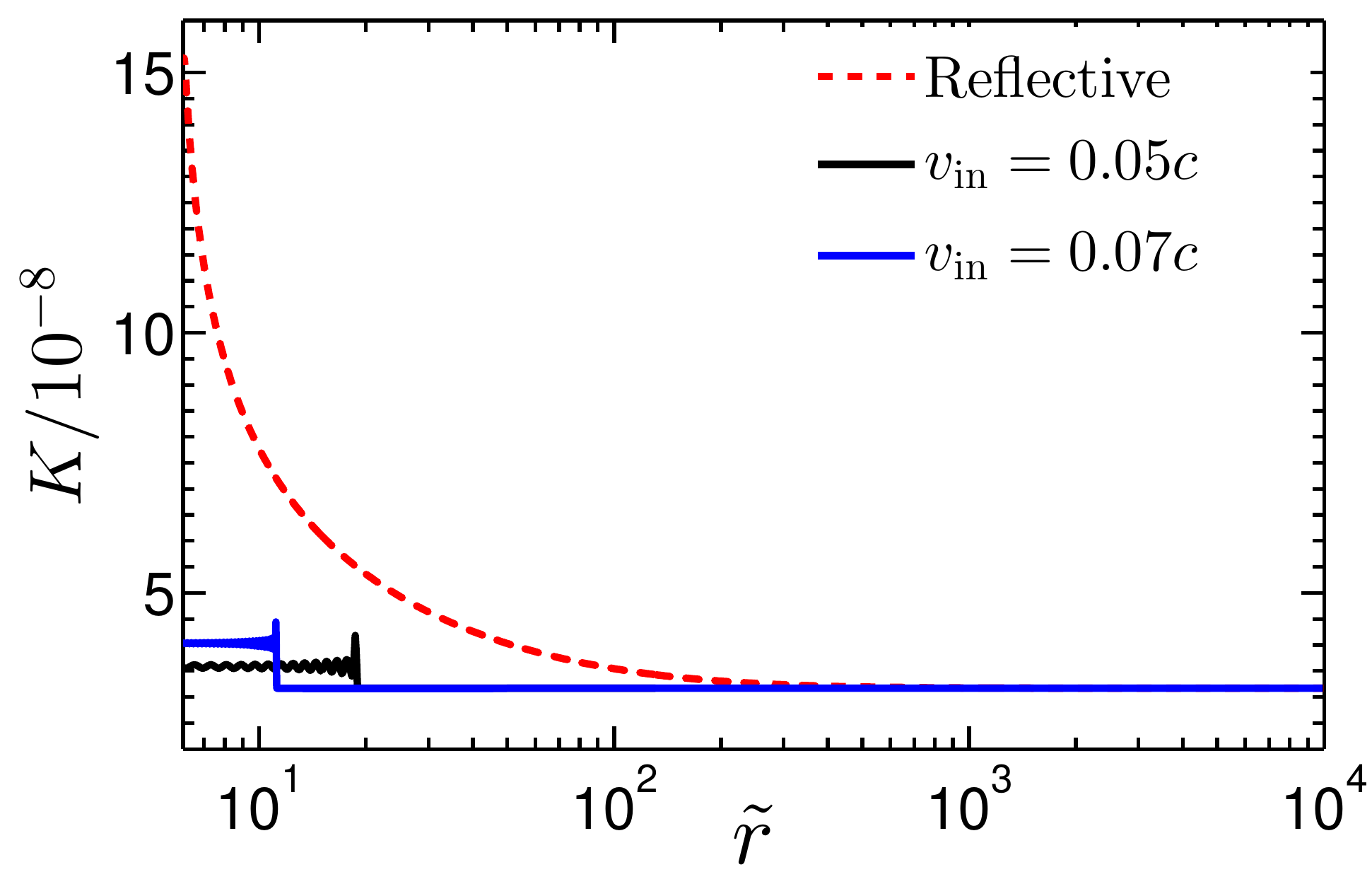}
	\caption{Entropy ($K \equiv p/\rho^{\gamma}$) profile in steady state from 1-D NS simulations with reflective and two different steady-shock inner boundary conditions ($v_{\rm in} = 0.05c$ and $v_{\rm in} = 0.07c$).}
	\label{fig:entropy}
\end{figure}

Fig. \ref{fig:entropy} shows the entropy (loosely defined as $p/\rho^\gamma$) profiles for the warm NS runs with reflective and steady-shock boundary 
conditions. For the reflective boundary condition run, entropy decreases smoothly with increasing radius. As discussed earlier, for the reflective boundary 
condition a shock is launched outwards, and it weakens and eventually vanishes, giving rise to a hydrostatic atmosphere in steady state. The entropy 
profile has the imprint of the shock strength; matter that crosses the shock earlier experiences a stronger shock and hence the entropy at inner radii is larger.
Entropy increases only when the fluid elements cross the shock; it is constant at all other times even in changing ambient conditions. Since 
entropy increases toward the direction of gravity in steady state, the flow is expected to be unstable to Schwarzschild convection (for applications to 
supernovae see \citealt{Herant1994,Scheck2008}). Since convection involves rise of underdense (hot) blobs relative to the background, it is essentially 
a 2-D phenomenon. Therefore, we carry out 2-D simulations, which are described in section \ref{sect:2-D}.

Fig. \ref{fig:entropy} also shows the entropy profiles for 1-D NS simulations with steady-shock boundary conditions. The flow in these cases is qualitatively different 
from the hydrostatic atmosphere obtained with the inner reflective boundary condition. For the steady-shock boundary condition, entropy is constant both outside 
and inside the shock, with a jump in entropy at shock-crossing. Small oscillations are seen in the post-shock entropy profiles. These are small amplitude 
entropy modes (in hydrostatic balance) arising because of reflections at the inner boundary. The standing shock is stable to radial 
perturbations (e.g., see \citealt{Blondin2003}).

An important question that arises is whether our steady NS profiles are described by the classic Bondi solutions. This question is addressed in detail
in section \ref{sect:bondi_regimes}. Here we just mention that none of the branches of the famous Mach-number versus radius plot of Bondi (c.f. top-left
panel of Fig. \ref{fig:cartoon}) can describe the
solution that we obtain in the steady of 1-D NS runs with reflective inner boundary condition. The reason for this is clear from Fig. \ref{fig:entropy} because
entropy ($p/\rho^{\gamma}$) is not constant as a function of radius. Bondi solution assumes a polytropic relation between pressure and density; i.e., 
$p/\rho^\Gamma$ is constant for {\em some} $\Gamma$. The entropy profile shown by a red dashed line in Fig. \ref{fig:entropy} does not satisfy $p/\rho^\Gamma=$
constant {\em for any $\Gamma$}. On the other hand, the solutions for a steady-shock inner boundary condition are isentropic inside and outside the shock 
and hence are described by various branches of the Bondi solution.

\subsubsection{2-D}
\label{sect:2-D}
\begin{figure*}
	\includegraphics[scale=0.92]{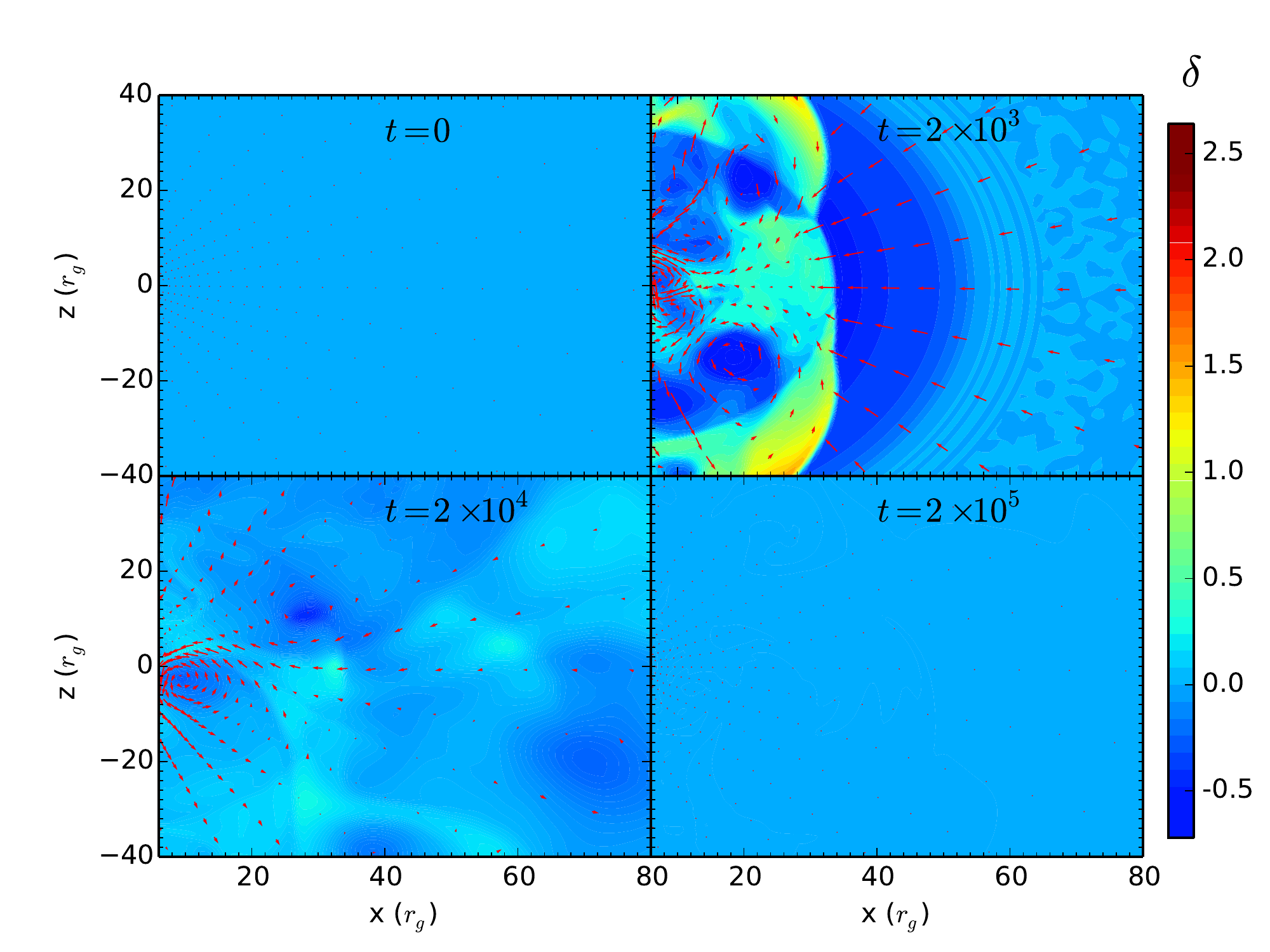}
	\caption{Snapshots of overdensity ($\delta (r,\theta) \equiv [\rho(r,\theta) - \bar{\rho}(r)]/\bar{\rho}(r)$; where $\bar{\rho}(r)$ is the $\theta-$averaged density) at different times from the 2-D simulation of NS with reflective inner boundary condition. Arrows show the flow velocity; local velocity is along the direction of arrows and the length of the arrow is proportional to the magnitude of velocity. For clarity, the velocity arrows are not placed at all grid points.}
		\label{fig:2-D_ref}
\end{figure*}

The results from 2-D BH simulations (with outflow inner boundary condition) match the 1-D runs, and in steady state (which is attained eventually for a warm ISM)
are well-described by the standard Bondi solution. Therefore, in this section we only discuss the 2-D NS simulations for a warm ISM, with both reflective and 
steady-state inner boundary conditions. These 2-D simulations are qualitatively different from 1-D.

\begin{figure}
	\includegraphics[scale=0.4]{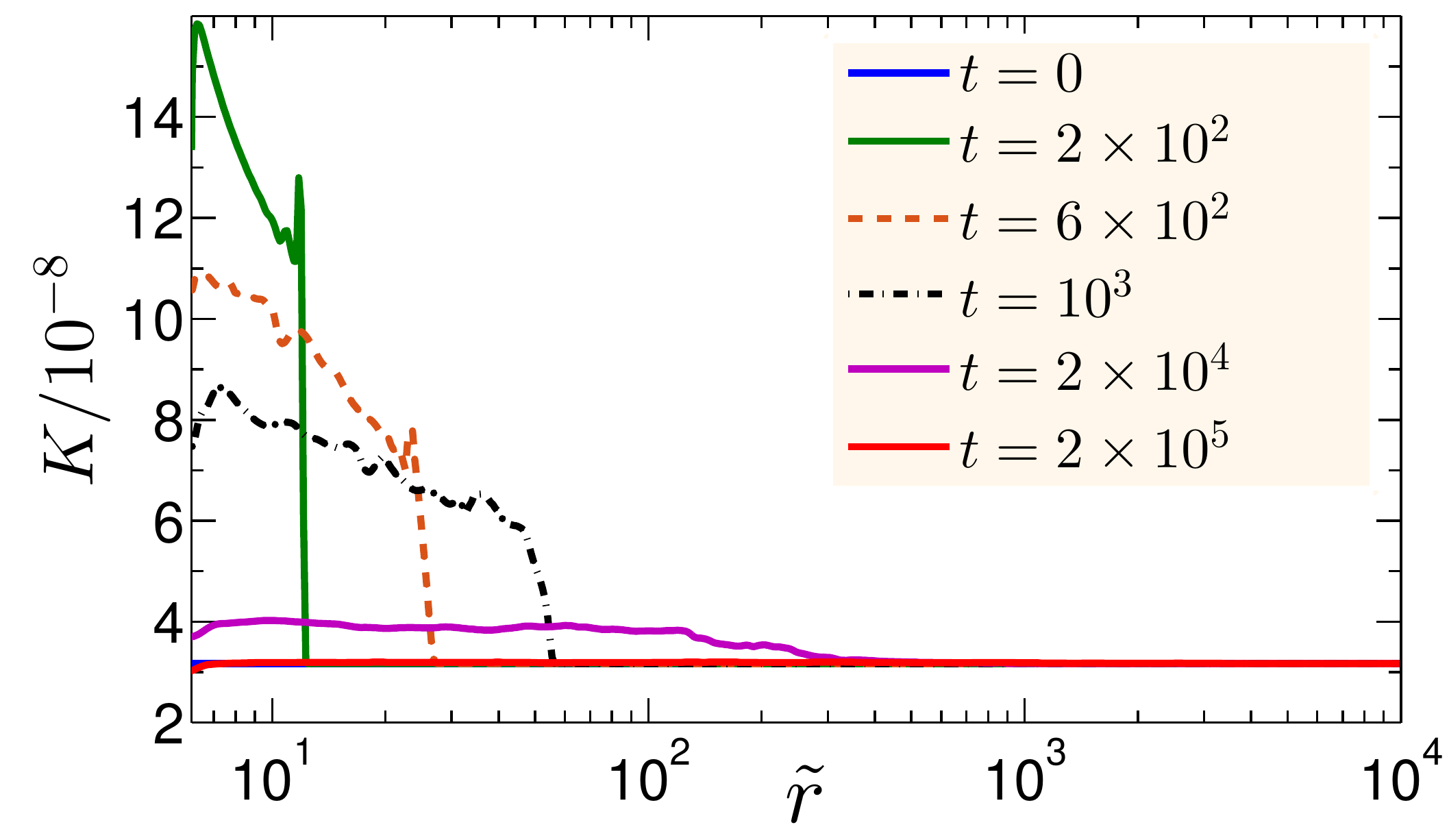}
	\caption{Angle-averaged entropy ($K \equiv \bar{p}/\bar{\rho}^\gamma$; $\bar{p}$ and $\bar{\rho}$ are $\theta$-averaged pressure and density) 
	profiles at different times for the 2-D NS simulation with the reflective inner boundary condition.}
		\label{fig:2-D_ent}
\end{figure}

{\em Reflective inner boundary condition:} Fig. \ref{fig:2-D_ref} shows the snapshots of overdensity ($\delta \equiv [\rho(r,\theta)-\bar{\rho}(r)]/\bar{\rho}(r)$; 
where, $\rho(r,\theta)$ is the density at $(r,\theta)$, $\bar{\rho}(r)$ is the $\theta-$averaged density at $r$) superposed with velocity arrows at different times for
the 2-D NS run with reflective inner boundary conditions. Initially ($t=0$) the density is uniform with $\delta = 0$ and the velocity vanishes. With time, NS starts 
accreting matter. Matter with supersonic velocity hits the NS surface and forms a shock which propagates outward in time. 
The top-right snapshot at $t=2\times10^3$ in  Fig. \ref{fig:2-D_ref} shows that the flow is radial before it is shocked, but within the shock spherical 
symmetry is broken. Also the shock location (most easily located by the sudden change in arrows representing velocities) is not spherically symmetric; at 
$t=2\times 10^3$ (top-right panel) the shock extends much further out along poles as compared to the equator. Fig. \ref{fig:entropy} shows that entropy is 
maximum toward the center for NS simulations with reflective inner boundary condition. Such an atmosphere is convectively unstable, and indeed the gas 
within the shock in Fig. \ref{fig:2-D_ref} shows convective swirling motions, with underdense ($\delta <0$) blobs rising and overdense ($\delta > 0$) blobs 
sinking with respect to the background  gas.
Snapshot at $t=2\times10^4$ (bottom-left panel) in  Fig. \ref{fig:2-D_ref}  
shows that convection (as measured by $\delta$) becomes weaker with time. As in 1-D (see the bottom panel of Fig. \ref{fig:den_NS_hot_cold} 
and red dashed line in Fig. \ref{fig:entropy}), the outer shock also becomes weaker with time in 2-D.
The snapshot at $t=2\times10^5$ (bottom-right panel in  Fig. \ref{fig:2-D_ent}) shows that after a sufficiently long time (longer than the sound crossing time
across the Bondi radius) the density fluctuations and velocities become negligible. By this time the outer shock vanishes and the radial entropy gradient is erased 
by convection. In this steady state the system is well described as a polytrope ($p/\rho^\gamma \approx$ constant) in hydrostatic equillibrium. In steady state
the mass accretion rate vanishes (because the velocity at the inner boundary is set to zero for a reflective inner boundary), like in 1-D. However, the key difference
from 1-D is that entropy ($p/\rho^\gamma$) is approximately constant (red-dashed line  in Fig. \ref{fig:entropy} is the entropy profile in 1-D).

Fig. \ref{fig:2-D_ent} shows the $\theta-$averaged entropy distribution as a function of radius at different times. Post-shock gas has a higher entropy because of
the entropy generated at the shock. With time the shock becomes weaker, with a lower post-shock entropy. The peak entropy at early times ($t=200$) in Fig. 
\ref{fig:2-D_ent} is similar to the entropy at the smallest radii for the 1-D run shown by red dashed line in Fig. \ref{fig:entropy}. However, with time the entropy
peak at the center flattens because of convection; higher entropy, underdense blobs rise leaving behind lower entropy gas at the center. In 2-D, not only does 
the shock front become weaker and moves out with time as in 1-D, the entropy profile within the shock is flatter. Eventually, at $t=2\times 10^5$ the entropy 
profile is perfectly flat and convection (and associated density and velocity perturbations) turns off.

\begin{figure*}
	\includegraphics[scale=.6]{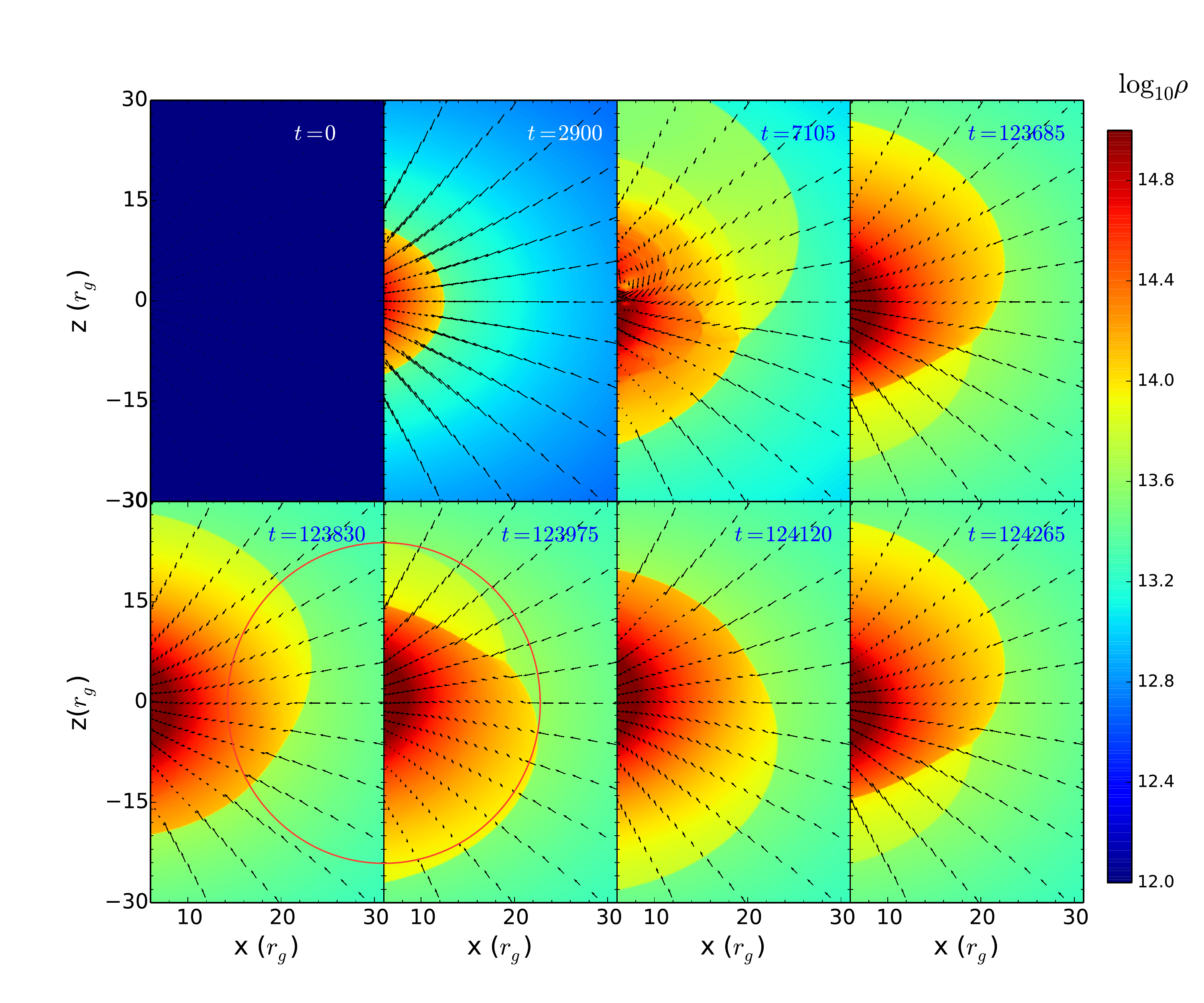}
	\caption{Density snapshots at different times for the 2-D 
	NS simulation  
	with steady-shock boundary condition. Arrows show the flow velocity. Top right and lower panels clearly show the 
	vertically oscillating ($l=1$ mode) 
	standing shock. The first two panels show the shock developing and moving outward in time. There are signatures of convection
	in the top-third panel at $t=7105~r_g/c$. The vertical shock oscillation period is roughly $580~r_g/c$ (see Fig. \ref{fig:vrvt_osc}). One can 
	also see the spherical higher frequency ($T=290~r_g/c$) mode outside the shock in the density snapshots at 
	$t=123685,~123830,~123975~r_g/c$.	 } 
	\label{fig:2-D_steady}
\end{figure*}

{\em Steady-shock inner boundary condition:} Fig. \ref{fig:2-D_steady} shows density snapshots and velocity vectors at various times for the 2-D NS 
simulation with the  steady-shock inner boundary condition ($v_{\rm in}=0.05c$). The initial evolution of the flow is similar to the reflective inner boundary condition run discussed above.  
Here too the shock propagates out in time as shown in the snapshots at $t=2900 r_g/c$ and $t=7105 r_g/c$ in Fig. \ref{fig:2-D_steady}. 
Convection is present only in the initial phase of the flow evolution. Unlike with reflective inner boundary, the shock does not disappear at late times; instead, 
a standing shock is formed, which is aspherical in shape and oscillates. 
Snapshots from $t=123685$ $r_g/c$ to $t=124265$ $r_g/c$ in Fig. \ref{fig:2-D_steady} show
the density oscillations in the (quasi)steady state. Two modes of oscillations can be seen in Fig. \ref{fig:2-D_steady}. First, the $l=1$ mode in which
the outer shock oscillates vertically. Snapshots from $t=123685$ $r_g/c$ to $t=124265$ $r_g/c$ represent a single oscillation period 
($T \approx 580~r_g/c$; the same period as $v_\theta$ oscillations in Fig. \ref{fig:vrvt_osc}) for this mode.
Another mode, which appears as an extra density feature beyond the shock at $t=123685,~123975,~124265$ $r_g/c$ is a radial compression 
mode that is hidden behind the $l=1$ mode at $t=123830,~124120$ $r_g/c$. This compression mode has half the time period of the $l=1$ mode
($T \approx 290~r_g/c$; oscillations with the same period are seen for the radial velocity in Fig. \ref{fig:vrvt_osc}) and is chiefly responsible for density 
and luminosity (c.f. Fig. \ref{fig:qpos}) fluctuations.

\begin{figure}
	\includegraphics[scale=0.48]{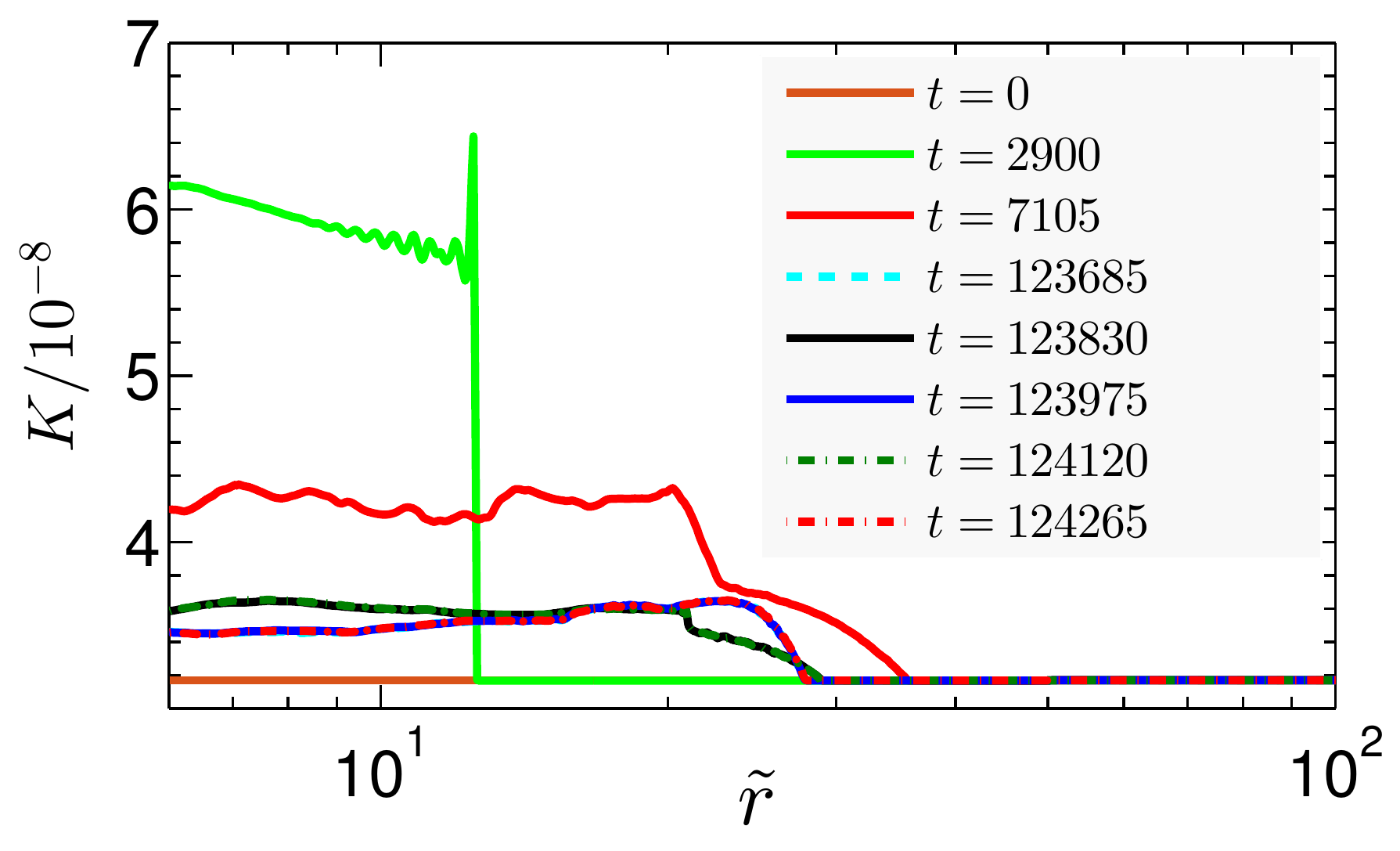}
	\caption{ Angle-averaged entropy ($K \equiv \bar{p}/\bar{\rho}^\gamma$; $\bar{p}$ and $\bar{\rho}$ are $\theta$-averaged pressure and density) 
	profiles at different times for the 2-D NS simulation with the steady-shock inner boundary condition. At late times the average shock radius remains
	stationary between $20-30 r_g$ (comparable to the stationary shock in 1-D; see the solid black line in Fig. \ref{fig:entropy}).} 
	\label{fig:2-D_steady_ent}
\end{figure}

Fig. \ref{fig:2-D_steady_ent} shows the angle-averaged entropy profiles for the 2-D NS simulation with the inner steady-shock boundary condition ($v_{\rm in}=0.05 c$). 
Initially (at $t=2.9 \times 10^3 r_g/c$) the entropy gradient is negative in the post-shock region. Eventually convection and advection through the inner boundary 
remove the entropy gradient and the 
post-shock entropy profile becomes almost flat. Therefore, as with the reflective inner boundary condition, convection does not play a significant role {\em after} 
the flow attains a steady state (as also noted by \citealt{Blondin2003}). Note that the angle-averaged post-shock entropy profiles are not perfectly flat and 
the shock location is smeared because the 
shock is not spherically symmetric. A comparison of the angle-averaged entropy profile in Fig. \ref{fig:2-D_steady_ent} with the 1-D profiles 
in Fig. \ref{fig:entropy}  is instructive. The shock location and the
entropy value for the 2-D steady-shock boundary condition run in steady state are similar to the corresponding values in the 1-D run.  Unlike with the reflective 
inner boundary condition, convection is not necessarily required to attain an isentropic post-shock steady state; even the analogous 1-D run (see black solid line 
in Fig. \ref{fig:entropy}) attains a constant entropy in the post-shock region because of advection of entropy out of the computational domain at the inner boundary.
Since energy is advected through the inner boundary (albeit subsonically) for the steady-shock inner boundary condition, the shock does not propagate to as large 
a distance as with the reflective inner boundary condition (compare Figs. \ref{fig:2-D_steady_ent} and \ref{fig:2-D_ent}).

\begin{figure}
	\includegraphics[scale=0.41]{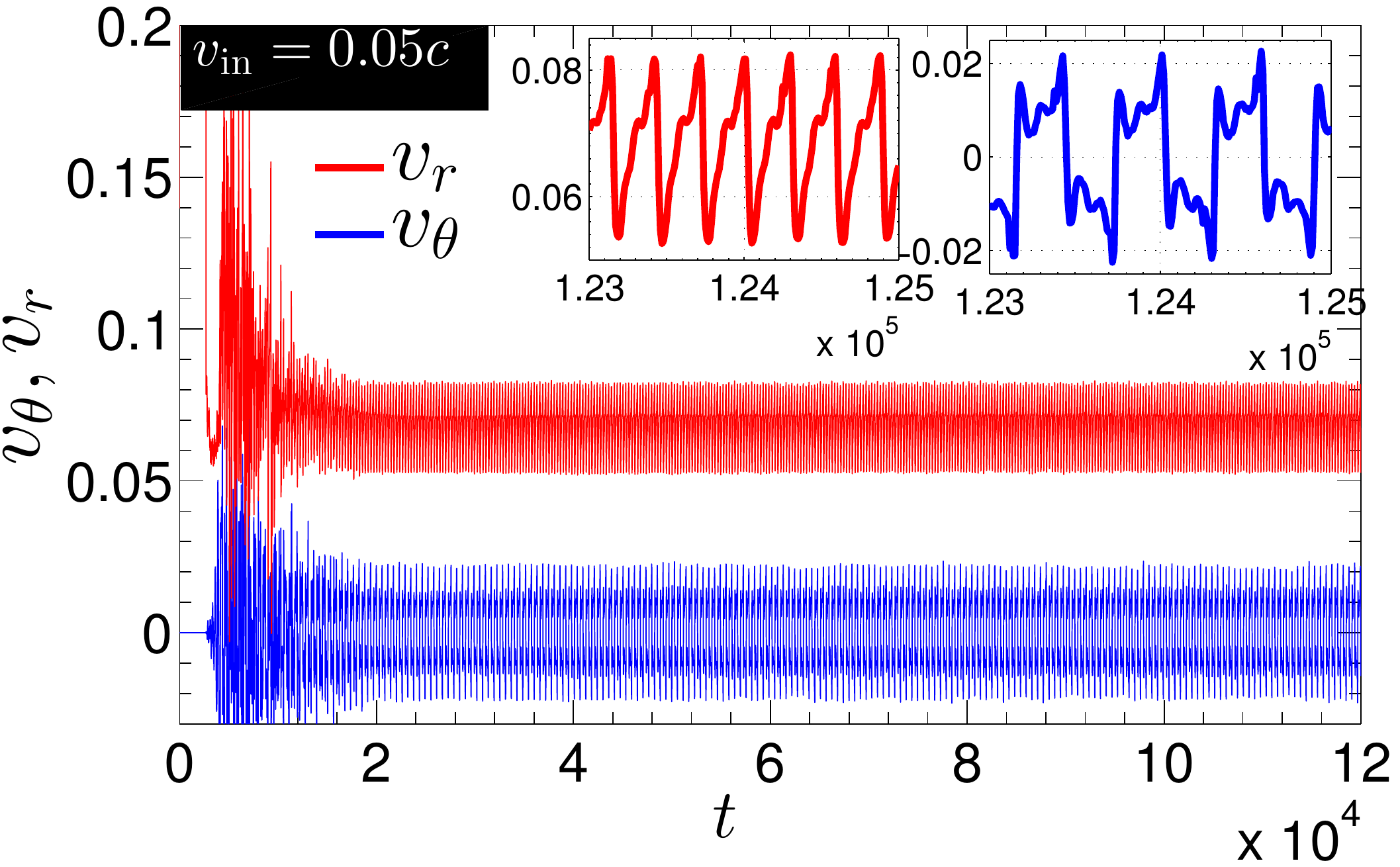}
	\caption{The value of radial and meridional velocities ($v_r,~v_\theta$) at a given point inside the shock ($r=8.1r_g,~\theta=\pi/2$) as a function 
	of time for the 2-D NS simulation with steady-shock inner boundary condition ($v_{\rm in}=0.05 c$). The radial velocity grows (with a negative sign) 
	at early times. After this, the oscillations grow and overshoot, and eventually settle down to a very coherent state. 
	The insets show the zoomed-in view 
	of oscillations at late times. While $v_r$ oscillates predominantly with a period of $290 r_g/c$, $v_\theta$ oscillates with a period of double this value 
	($580 r_g/c$; this matches global density oscillation frequency in Fig. \ref{fig:2-D_steady}). Note that $v_r$ has a non-zero mean, 
	corresponding to the inward radial advection velocity.}
	\label{fig:vrvt_osc}
\end{figure}

The large amplitude global oscillations seen in Fig. \ref{fig:2-D_steady} result from the vortical-acoustic instability of standing shocks 
known as the standing accretion shock 
instability or {\em SASI} (\citealt{Foglizzo2000,Foglizzo2007,Foglizzo2012}). The instability is thought to arise due to the unstable 
advection-acoustic cycle in which vorticity/entropy perturbations are advected inwards and sound waves propagate outwards due to reflection 
at the inner boundary and the shock, respectively (e.g., see left panel in Fig. 1 of \citealt{Guilet2012}). 

\begin{figure}
	\includegraphics[scale=0.43]{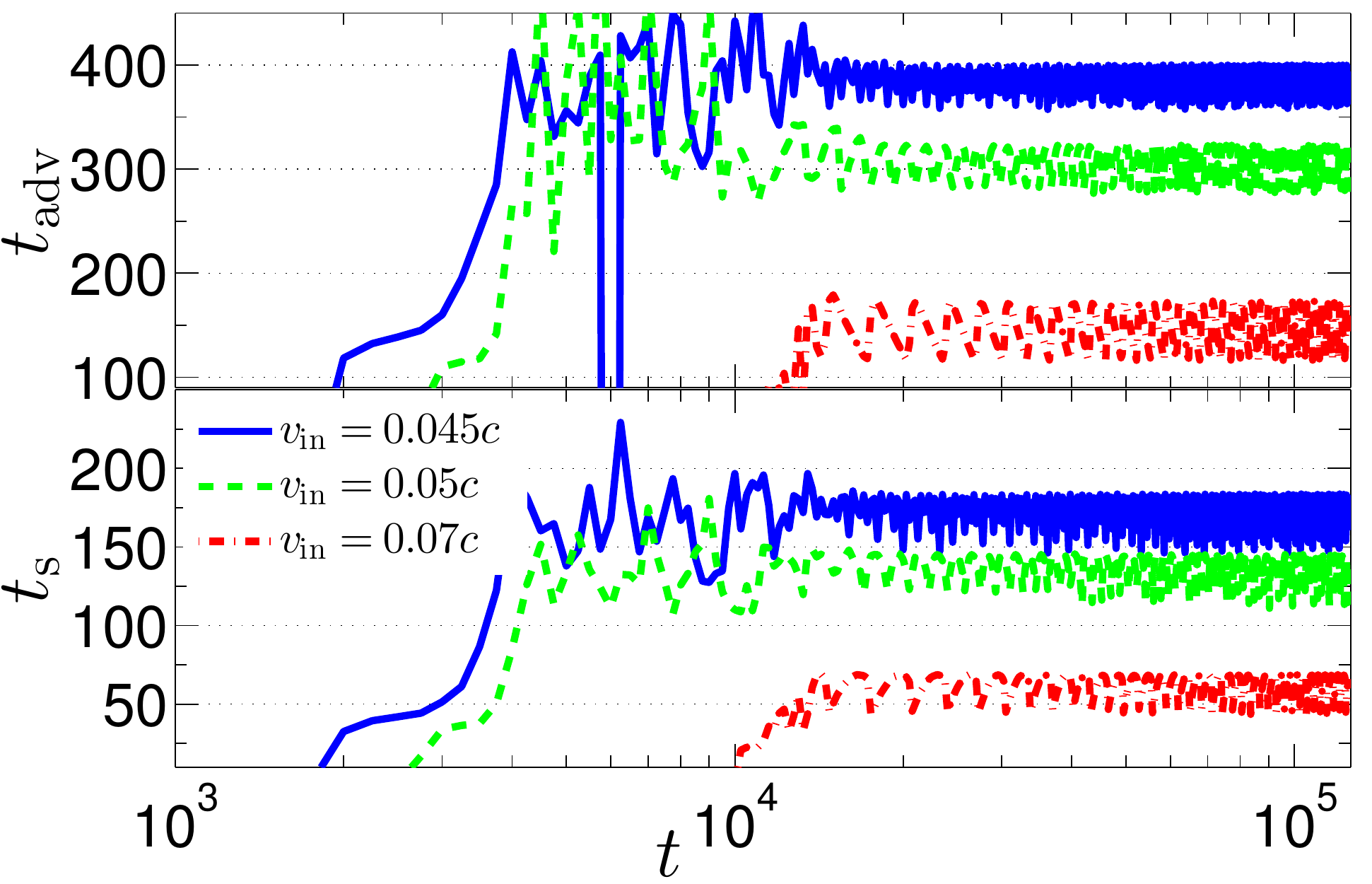}
	\caption{Radial advection time ($t_{\rm adv} \equiv  \int_{r_{\rm in}}^{r_{\rm sh}} dr/\bar{v}_r$; $\bar{v}_r$ is the mass-weighted, 
	angle-averaged radial velocity as a function of $r$; $r_{\rm sh}$ is the shock radius which changes in time; Table \ref{tab:tab1} states the value of 
	average and maximum shock radius) 
	and radial sound crossing time 
	($t_s \equiv \int_{r_{\rm in}}^{r_{\rm sh}} dr/\bar{c}_s$; $\bar{c}_s$ is the mass-weighted, 
	angle-averaged sound speed as a function of $r$) as a function of time for three NS steady-shock inner boundary condition runs with 
	$v_{\rm in}=0.045,~0.05,~0.07 c$. For a smaller inner velocity the shock location is farther out, and the advection and sound crossing 
	times are longer. Also, in steady state, the advection time is longer than the sound crossing time as the flow within the shock is subsonic.}
	\label{fig:adv_snd_time}
\end{figure}

\begin{table}

\caption{Results from 2-D SASI simulations in steady state}
\begin{tabular}{c c c c c c c}
\hline
$v_{\rm in}/c$ & $r_{\rm sh} $ & $r_{\rm sh, max}$ & $t_{\rm adv}^\star$ & $t_{\rm s}^\star$ & $T(v_r)$ & $T(v_\theta)$ \\
 		   &  		(1-D)		   &  (2-D)			     &   $(r_g/c)$ & $(r_g/c)$ & $(r_g/c)$  & $(r_g/c)$  \\
\hline
0.045 & 26.7 & 33.1 & 392 & 163  & 400 & 800 \\
0.05 & 20.8 & 28.8   & 291 &  145 & 290 & 580 \\
0.07 & 11.8 & 17.6    & 162  &   58  & 120 & 240 \\
\hline
\end{tabular}
$^\star$ time-averaged in steady state (see the caption of Fig. \ref{fig:adv_snd_time}).
\label{tab:tab1}
\end{table}

Fig. \ref{fig:vrvt_osc} shows the variation
of radial and meridional velocities at a fixed point within the oscillating shock. There are two prominent oscillation time periods, roughly $290 r_g/c$ 
(clearly seen in $v_r$ oscillations) and $580 r_g/c$ (seen in $v_\theta$ oscillations; also prominent in global density oscillations of Fig. \ref{fig:2-D_steady}).
To study the effects of the velocity imposed at the inner boundary, we have run NS simulations using 
$v_{\rm in}=0.45,~0.07 c$. A smaller velocity at the inner boundary pushes the steady shock outwards, and we expect the radial advection time and hence the SASI 
oscillation period to be longer. Fig. \ref{fig:adv_snd_time} shows the time variation of advection and sound-crossing times across the shock for different
$v_{\rm in}$. As expected, the timescales are longer for a smaller $v_{\rm in}$. Table \ref{tab:tab1} lists various important quantities 
(shock location from 1-D simulations, maximum shock location in 2-D, time-averaged advection and sound-crossing times from the shock to the inner boundary, and the time periods for $v_r$ and $v_\theta$ oscillations) from our 2-D steady shock NS simulations with three different $v_{\rm in}$. 
In all cases, the period of $v_\theta$ oscillations 
(which coincides with global $l=1$ density oscillations seen in Fig. \ref{fig:2-D_steady})
is double that of radial velocity oscillations (which oscillates at the same frequency as the radial mode seen in Fig. \ref{fig:2-D_steady}). 
The time period of the radial velocity oscillation roughly matches the radial advection time for $v_{\rm in}=0.045c$ and $v_{\rm in}=0.05c$, 
but deviates for $v_{\rm in}=0.07c$. This deviation may be due the smallness of the size of the post-shock region for $v_{\rm in}=0.07c$. 
There is still no precise prediction for the oscillation period of SASI, but the advection timescale, which shows a similar 
trend as the measured shortest oscillation period in all cases, gives a good estimate (\citealt{Guilet2012}).

\section{Discussion and implications}
\label{sect:discussion}

In this section we discuss the astrophysical implications of our results. Although our set up is quite idealized, we can apply some of our results to
interpret various observations of radiatively inefficient accretion on to compact objects.

\subsection{Three regimes of adiabatic spherical accretion}
\label{sect:bondi_regimes}

\begin{figure*}
        \includegraphics[scale=0.8]{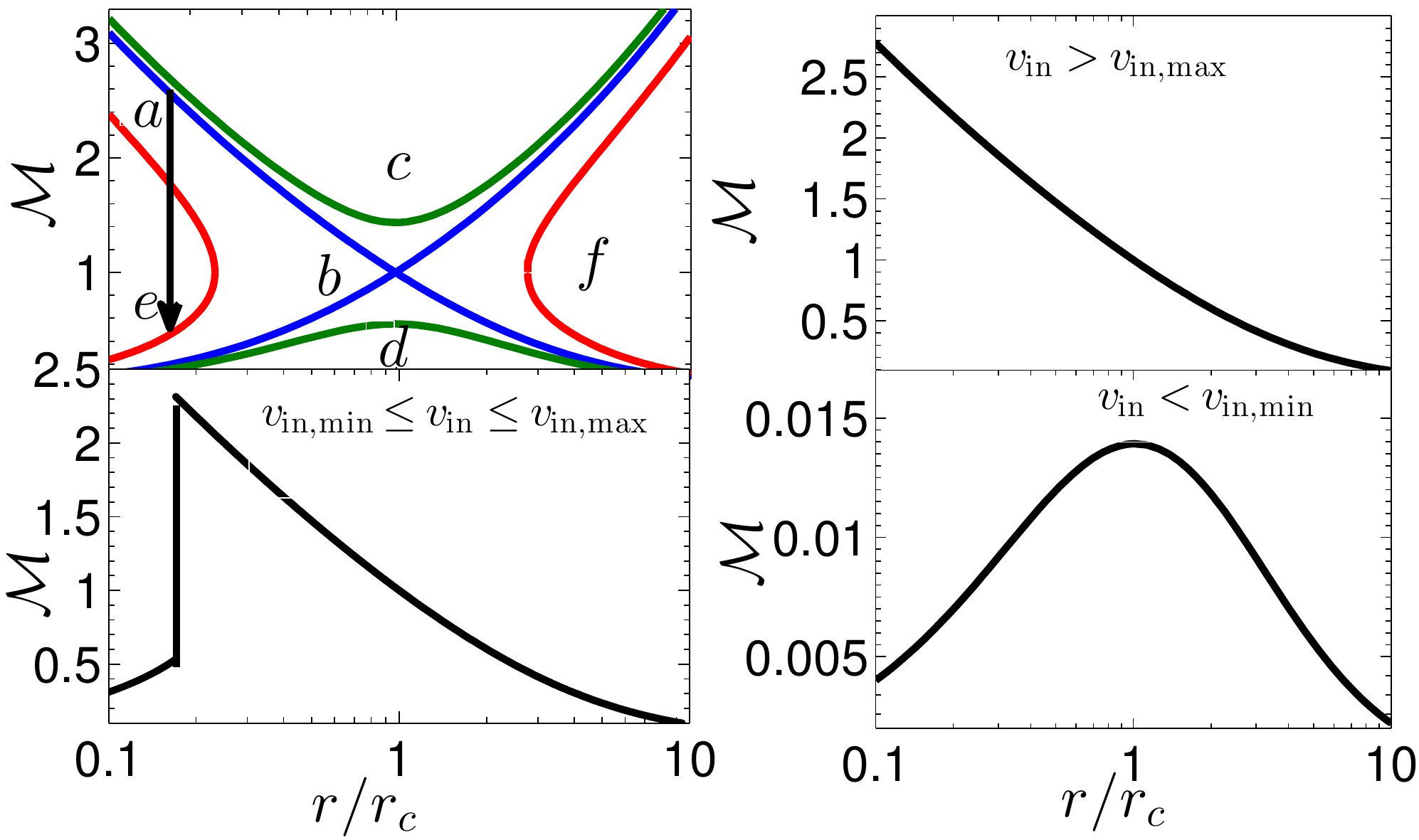}
        \caption{Mach number as a function of the scaled radius for spherical accretion in 1-D. Top-left panel shows all the possible branches for 
        steady, isentropic accretion. The other panels show the different flow regimes realized for the different values of
        infall velocity $v_{\rm in}$ at the inner boundary ($r_{\rm in}$).}
                \label{fig:cartoon}
\end{figure*}

From our results in sections \ref{sect:similarity} and \ref{sect:simulation} it is clear that spherical accretion, in absence of angular momentum, admits
three solutions consistent with the inner and outer boundary conditions for BHs and NSs. For BHs, which allow supersonic advection of matter at 
the inner boundary, accretion is described by the classic transonic solution. For NSs, in which a surface stops the infalling matter, there 
are two possibilities: (i) isentropic, 
throughout subsonic accretion with an accretion rate smaller than the Bondi rate (in 2-D the subsonic flow becomes 
isentropic because of convection; see section \ref{sect:2-D}); 
(ii) a solution with a steady shock in 1-D with an accretion rate
equal to the Bondi value (as we note in section \ref{sect:2-D}, this solution is unstable to SASI which results in global oscillations in 2-D; 
the average accretion rate still equals the Bondi value though).

Fig. \ref{fig:cartoon} shows the variation of Mach number (${\cal M} \equiv v/c_s$) as a function of radius scaled to the sonic radius in 1-D ($r_c$). 
The top-left
panel shows all branches of the solution (\citealt{Holzer1970}). The transonic branch (a) is the classic accretion solution, which is also shown in the 
top right panel. The bottom-left panel shows a solution with a steady shock (this shock is unstable to SASI in 2-D and 3-D). 
For a small enough inflow velocity at the inner radius, the solution
is the subsonic settling/sinking atmosphere shown in the bottom-right panel (also known as the breeze solution; it is branch (d) in the top-left panel).

\begin{figure}
        \includegraphics[scale=0.4]{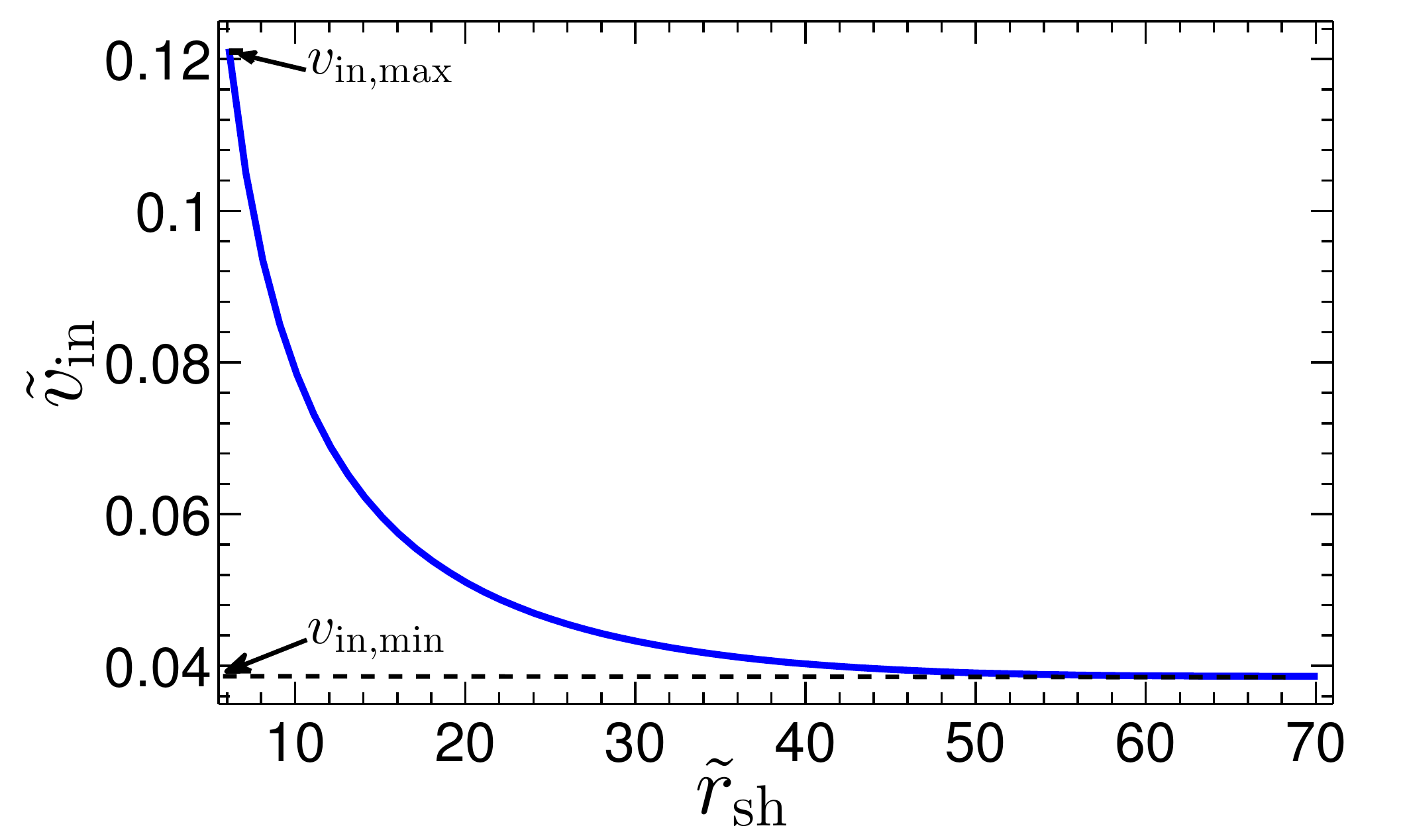}
        \caption{The inflow velocity at the surface of NS (i.e., the inner boundary) $v_{\rm in}$ (in units of $c$) as a function of the location of the 
        steady shock $r_{\rm sh}$ (in units of $r_g \equiv GM/c^2$). The same choice of parameters is made as the rest of the paper. The steady-shock 
        solution is realized only if $v_{\rm in, min} \leq v_{\rm in} \leq v_{\rm in, max}$.}
        \label{fig:vel_shock}
\end{figure}

The steady-shock solution connects the transonic accretion branch (a) to the higher entropy branch (e) with the same accretion 
rate (\citealt{McCrea1956}). Other Rankine-Hugoniot jump conditions must also, of course, be satisfied. Branches (a) and (d), however, cannot be 
connected by a shock because (d) has a lower entropy for the same accretion rate. For obtaining the steady-shock solution we solve the 
steady state forms of Eqs. (\ref{eq:mass}) and (\ref{eq:momentum}) in 1-D (assuming spherical symmetry) with  the polytropic equation 
of state $p=K\rho^{\gamma}$. We apply the Rankine-Hugoniot shock conditions at the chosen shock location $r_{\rm sh}$ (which must lie between the 
stellar surface and the sonic point). For a given shock location, the post-shock flow traces out a unique trajectory in the ${\cal M}-r/r_c$ plot, 
with a non-zero  (subsonic) velocity at the NS surface. The shock location is very sensitive to the value of velocity at the inner boundary, $v_{\rm in}$. 

Fig. \ref{fig:vel_shock} shows the variation of the velocity at the inner boundary (at $6 r_g$) as a function of the shock location for
our choice of parameters (e.g., $c_{s \infty}$). The maximum value of the inner velocity ($v_{\rm in,max}$) occurs when the shock forms just at
the surface and the minimum value occurs when the shock occurs just inside the sonic point ($\approx 71 r_g$ for our choice of parameters).
The steady-shock solution is not allowed for the inner velocity outside this range. An isentropic, subsonic settling solution occurs for 
$v_{\rm in} < v_{\rm in, min}$ and a transonic solution (similar to that for a BH) occurs for $v_{\rm in} > v_{\rm in, max}$.
It should be noted that mass accretion rates for both 
the transonic solutions (with and without shocks) are $\dot{M}=\dot{M}_B$, where $\dot{M}_B$ is the Bondi accretion rate. But, for the settling solution
the accretion rate can be much smaller than the Bondi value ($\dot{M} \ll \dot{M}_B$).\\

\subsection{Cooling and $v_{\rm in}$}
\label{sect:vin}
Section \ref{sect:bondi_regimes} shows that adiabatic spherical accretion is very sensitive to the velocity at the inner boundary, $v_{\rm in}$.
A key question is how this settling velocity is determined at the surface of an accreting star. Various factors, such as the porosity of the accreting 
surface and cooling, govern the velocity at the inner boundary. We have carried out 1-D simulations with free-free 
cooling to assess its role in setting $v_{\rm in}$, and consequently, the accretion regime. We do not discuss these simulations
in detail here, but just mention the salient features. 

If radiative cooling in the dense, inner accretion flow is efficient 
(i.e., cooling time is shorter than the dynamical time), a cooling layer on top of the stellar surface 
can form (e.g., see \citealt{Blondin2003}). In steady state, even with a reflective inner boundary condition (valid for a NS), gas can accrete on to the cooling layer, 
with most of the accretion energy 
being radiated away (this corresponds to the $v_{\rm in} > v_{\rm in, max}$ transonic solution, except that energy cannot be advected but has to be radiated). 
If cooling is less efficient, even with reflective/non-absorbing inner boundary condition, we can realize the $v_{\rm in, min} < v_{\rm in} <  
v_{\rm in, max}$ regime in which a steady shock is formed. In this case, substantial energy may still be lost radiatively but thermal pressure 
is maintained by converting a fraction of accretion energy into heating.  For inefficient radiative cooling and reflective inner boundary condition, 
an outward propagating shock is expected. In steady state the shock weakens and leads to a hydrostatic atmosphere, similar to the
$v<v_{\rm in, min}$ scenario. Cooling increases the accretion rate for spherical BH simulations (which we model with inner outflow boundary conditions); most energy and mass are still advected on to the BH. 

Earlier authors have studied the effects of cooling on steady accretion shocks and their global oscillations (\citealt{Saxton2002}). These global
oscillations are analogous to the oscillations seen in some of our NS simulations, except that our oscillations are controlled by a simpler parameter, 
$v_{\rm in}$ (see section \ref{sect:bondi_regimes}). Unlike our non-radiative simulations in which the standing shock is unstable only in two 
and three dimensions, the shock is unstable even in 1-D with cooling (\citealt{Langer1981,Chevalier1982}). A detailed comparison with the simulations 
with cooling is left for future. 

\subsection{ Flow of mass and energy}

\begin{figure}
        \includegraphics[scale=0.4]{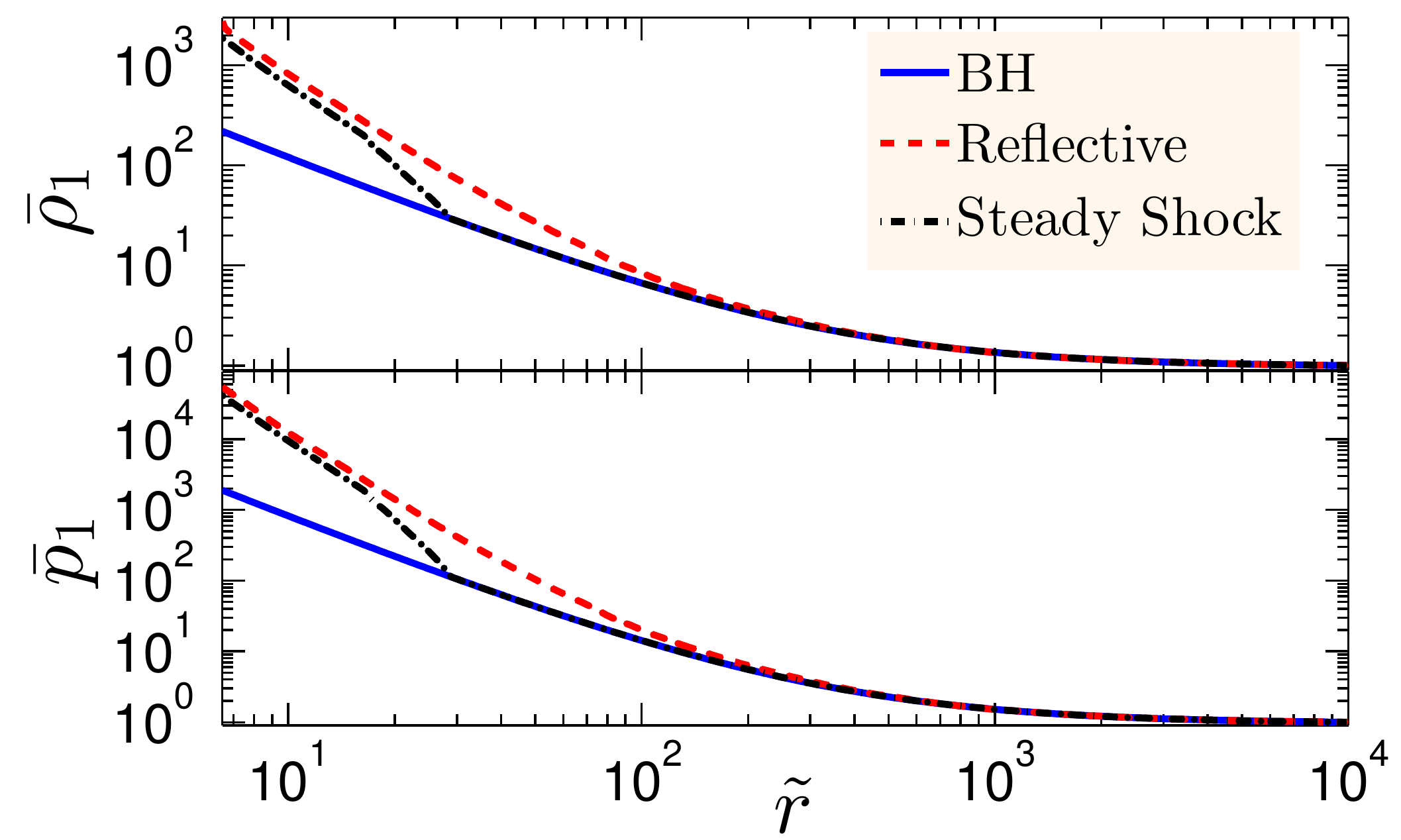}
        \caption{Normalized angle-averaged density ($\bar{\rho}_1 = \bar{\rho}/\rho_0$; top panel) and pressure ($\bar{p}_1 = \bar{p}/p_0$; bottom panel)
        profiles in the steady state of 2-D simulations as a function of the scaled radius (scaled to $r_g$; the sonic radius is 
        $r_c \approx 71 r_g$) for BH and NS simulations 
        (with both reflective and steady-shock inner boundary conditions).}  
        \label{fig:rhop2d}
\end{figure} 

Fig. \ref{fig:rhop2d} (top panel) shows the steady, angle-averaged density profiles for accretion on to a BH, and a NS with reflective and steady-shock 
inner boundary conditions. For 
both reflective and steady-shock boundary conditions, the density in the inner region of NS accretion flow is higher. For steady-shock boundary
condition, all profiles are identical to the Bondi/BH solution outside the shock location. Therefore, the accretion rate
in steady state is the same as the BH/Bondi accretion rate. Pressure and density are higher in the subsonic
inner region for a NS. Unlike a BH in which matter falls freely at inner radii, pressure balances most of the inward gravitational pull for a NS. 
The accretion
rate for the inner reflective boundary condition (and even for cases with $v_{\rm in} < v_{\rm in, min}$) is smaller than the Bondi value. In fact, in 
the steady state, accretion rate with reflective boundary condition should be identically zero.

Since NSs are less massive than BHs, in similar ambient conditions, the mass accretion rate for NSs should be smaller than the corresponding value for 
BHs. In $v_{\rm in} > v_{\rm in, min}$
regime, $\dot{M}=\dot{M}_B \propto M^2$, and even the scaled mass accretion rate $\dot{m} \equiv \dot{M}/\dot{M}_{\rm Edd} \propto M$ 
($\dot{M}_{\rm Edd} \propto M$ is the Eddington accretion rate) is larger for BHs 
as compared to NSs. For $v_{\rm in} < v_{\rm in, min}$ due to the NS surface, the accretion rate is further suppressed ($\dot{M} \propto v_{\rm in}$ for 
$v<v_{\rm, min}$). Therefore, in the radiatively inefficient accretion regime, the NS accretion rate can be orders of magnitude smaller than a BH.

It is instructive to study the flow of mass and energy in spherical accretion in different regimes. In the simplest adiabatic Bondi regime, gravitational
energy is converted to the kinetic energy of the infalling supersonic gas, which is advected on to the BH. In steady state, {\em all} the gravitational
power ($GM\dot{M}/2 r_{\rm in}$) is accreted by the BH. Even in presence of cooling, owing to the low density and temperature in the 
inner regions (see the top panel of Fig. \ref{fig:rhop2d}), radiative losses 
are subdominant for BHs.
Since energy cannot be advected through the inner boundary for NSs, {\em all} the gravitational power is radiated in this regime 
($G M \dot{M}/r_{\rm in}$, some upstream of the cooling layer and most within it). If matter cannot radiate efficiently close to the NS surface (likely
at low accretion rates), the accretion rate is suppressed by several 
orders of magnitude compared to the Bondi value. In this case, in steady state, the gravitational power extracted is small but all of it 
goes into radiative cooling.

We should mention here that any deviation from spherical symmetry (e.g., \citealt{Stone1999,Proga2003}), any chance of thermalization of energy 
(for example due to magnetic fields and their dissipation; e.g., \citealt{Igumenshchev2002}), etc. cause non-radiative accretion on to even BHs 
to be qualitatively different from the idealized Bondi solution, and in fact, closer to the NS solution. In this case, the mass accretion rate can be 
orders of
magnitude smaller than the Bondi estimate, as is the case for Sgr A* in the Galactic center (\citealt{Baganoff2003}). 
One key difference of our simulations from previous BH simulations with 
angular momentum and magnetic dissipation is that the midplane density is shallower in them (compare Fig. \ref{fig:rhop2d} with Fig. 5 in 
\citealt{Stone1999} and Fig. 13 in \citealt{Sharma2008}). The density profile in simulations with angular momentum depends on the form of 
accretion viscosity (Eq. 12 in \citealt{Das2013}). The profiles at inner radii in relativistic simulations (unlike us, \citealt{Sharma2008} use a 
pseudo-Newtonian potential to capture GR effects) are affected by the presence of the innermost stable circular orbit (ISCO), close 
to which the density profile flattens. Also, the density in the inner regions is flatter if entropy is larger at the center (e.g., compare the top 
panel of Fig. \ref{fig:rhop2d} with Fig. \ref{fig:steady}); this may be due to magnetic dissipation close to the center. This discrepancy in the 
density profile will be resolved in the future.

\subsection{QPOs and SASI}

\begin{figure}
        \includegraphics[scale=0.4]{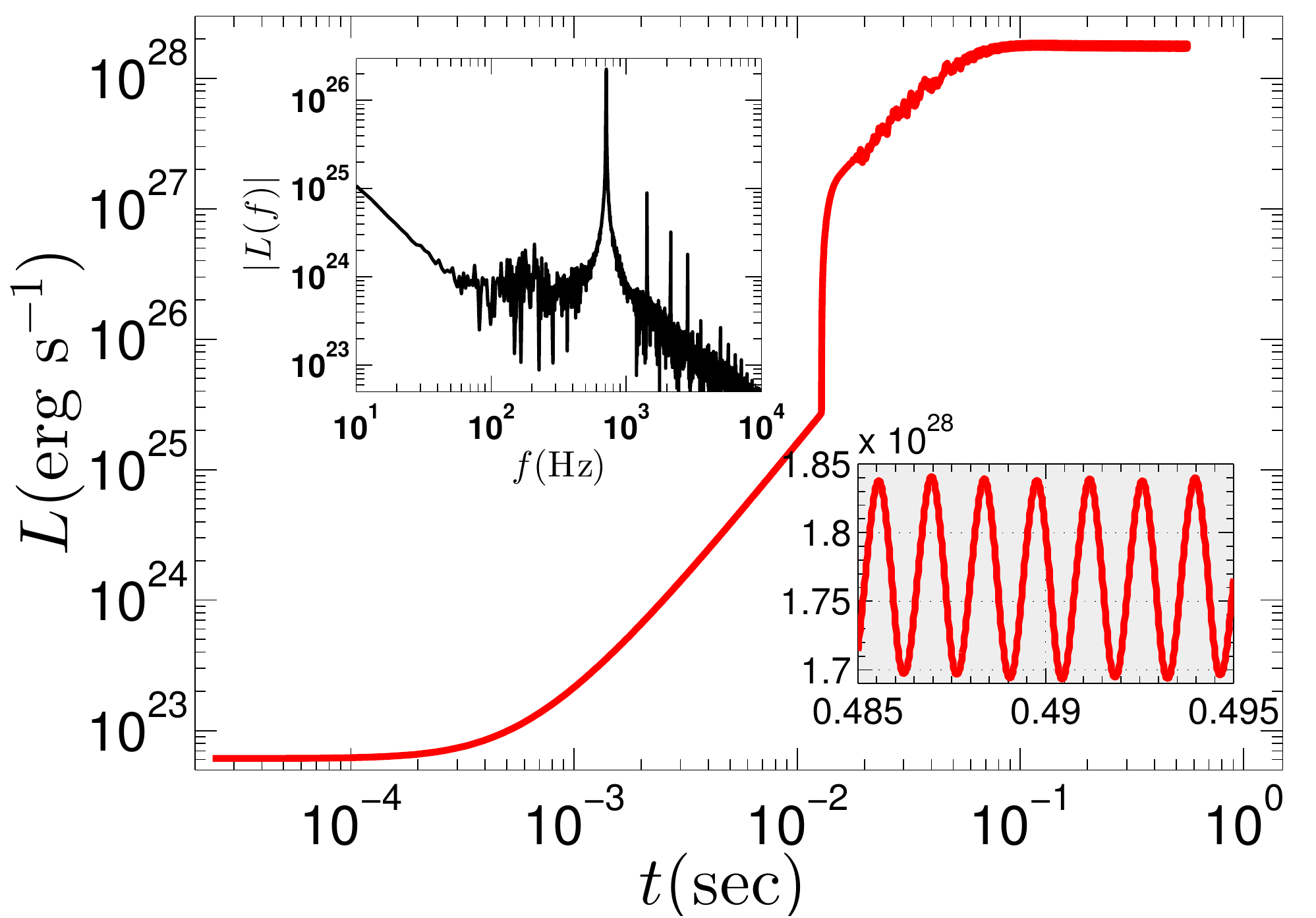}
        \caption{Free-free luminosity within $r_{\rm sh,max}$ ($\int_{r_{\rm in}}^{r_{\rm sh, max}} 1.4\times 10^{-27} [\rho/m_p]^2 T^{1/2} dV$; \citealt{Rybicki1986}) 
        as a function of time, scaled to a NS mass of 1 $M_\odot$; $\rho_\infty = 10^{12} m_p$ ($m_p$ is proton mass) gives $\dot{M}_B \approx 9 \times 10^{-13} 
        M_\odot {\rm yr}^{-1}$ (Eq. 2.36 in \citealt{Frank2002}; this is $\approx 4 \times 10^{-5}$ of the Eddington limit). 
        The bottom-right inset shows a zoomed-in version in steady state. The top-left inset shows a power spectrum with the most prominent peak
        at 713 Hz, which in dimensionless units, corresponds to a time period of $290 r_g/c$, which matches the $v_r$ oscillation frequency in Fig. 
        \ref{fig:vrvt_osc} and Table \ref{tab:tab1}. Peaks are also seen for higher harmonics. The power spectrum is taken in the steady state (from 
        0.1 to 0.56 s).}
        \label{fig:qpos}
\end{figure}

The presence of the standing accretion shock instability (SASI) for certain inner boundary conditions leads to the exciting possibility of explaining 
some of the quasi-periodic-oscillations (QPOs; e.g., see \citealt{Psaltis1999,Remillard2006,Mukhopadhyay2003,Mukhopadhyay2009}) observed in 
BH and NS XRBs accreting in the hard and 
intermediate spectral states. 

Typically the observed QPO frequency increases with the increasing mass accretion rate. This is expected in the SASI model 
because a larger accretion rate implies higher density, at which effective velocity at the inner radius ($v_{\rm in}$) increases and the shock 
moves in. A shorter advection time leads to a higher SASI/QPO frequency. In this model the hard and intermediate spectral states of BHs and NSs
(corresponding to $ v_{\rm in, min} \leq v_{\rm in} \leq v_{\rm in, max}$) should show QPOs. Recall that the accretion flow should suddenly
slow down in order for a steady shock to form. This slow down happens due to a hard surface in case of a NS and due to the formation of a 
centrifugal barrier in case of a BH. For non-radiative flows at very low accretion rates, the shock is too far away or it totally 
disappears. For very large accretion rates the shock is radiative, matter falls freely, and there is 
no resonant cavity formed.

Fig. \ref{fig:qpos} shows the luminosity due to free-free emission within the shocked accretion flow. While the shocked material in the simulations is 
very hot ($T \gtrsim 10^{11}$ K), the electrons in hot accretion flows are much cooler than the protons (e.g., \citealt{Sharma2007,Yuan2014}) and 
therefore lightcurves from simulations (which assume the same electron and proton temperatures) should be only taken as trends.
 As expected, luminosity increases
as the hot, dense post-shock gas is created. At late times the luminosity shows coherent oscillations (shown in the bottom inset). The top
inset shows the power spectrum of light-curve in steady state and shows peaks at 713 Hz and its higher harmonics. This frequency corresponds 
to a time period of $290 r_g/c$ (for $M=1 M_\odot$), seen for $v_r$ oscillations in Fig. \ref{fig:vrvt_osc}. The $v_\theta$ and $l=1$ density 
oscillations seen in Fig. \ref{fig:2-D_steady}
have double the time period (and half the frequency), but are not prominent in the power spectrum because density displacement does not substantially 
change the luminosity (radial compression, on the other hand, appreciably changes luminosity). The results are expressed assuming a NS mass of 
one solar mass, but can be scaled easily with the mass of the compact object. The shock location, and hence the inverse of the QPO frequency (identified 
with $v_r$ oscillations in Fig. \ref{fig:vrvt_osc}) is $\propto M$ for the same ambient conditions. We note that the $Q-$value for the oscillations is quite high,
but a slow modulation of $v_{\rm in}$ in time due to cooling can lower it such that it matches observations. Moreover, there will be contributions to the lightcurve
from the MHD turbulent accretion disk which can lower the Q-factor (\citealt{Reynolds2009}).

One of the most important X-ray transients, GRS$~1915+105$, is classified in twelve different temporal classes
in terms of its timing properties (lightcurves; \citealt{Belloni2000}). 
Interestingly, while some of the classes, e.g., $\chi$ exhibiting the low/hard 
state, show continuous jets (as measured from persistent radio emission), some others, e.g., $\theta$ corresponding to the
intermediate state, show episodic jets. \citet{Chakrabarti1999,Molteni1996,Ryu1997} argued that steady jet is produced by the
steady shock seen in their 2-D simulations. The steady jet may thus describe the $\chi$ state of GRS~1915+105. We also see a quasi-steady 
oscillating shock in our 2-D simulations if we fix the velocity at the inner boundary  to an intermediate value (section \ref{sect:2-D}).
We speculate on how a transient jet observed in the $\theta$ state may be produced. The spectral state transition from a low-hard ($\chi$)
to an intermediate state ($\theta$) happens due to an increase in the the mass accretion rate (and the resulting efficient cooling; \citealt{Das2013}).
Because of enhanced cooling, the effective velocity at the inner boundary increases beyond the maximum value for which a steady shock 
can form ($v_{\rm in, max}$; see Fig. \ref{fig:vel_shock}; see also section \ref{sect:vin}). In this scenario, the shock becomes stronger for a short time 
(of order the cooling time) and eventually disappears, plausibly leading to strong episodic jets as seen in the intermediate state. 
Quantitative comparisons with observations using realistic simulations is left for future. 

\subsection{Caveats: role of rotation, asymmetry, and magnetic fields}

Although for simplicity we have assumed spherical symmetry, and neglected rotation and magnetic fields, they are necessary 
ingredients for a realistic model of accretion.
While spherically symmetric accretion on to a BH does not admit a shock, steady state models of an accretion flow with angular momentum 
allows a shock. The shock may occur at radii where the pre-shock flow is accelerating or decelerating (e.g., see Fig. 6 in \citealt{chakrabarti1989}; 
the shock outside/inside the O-type sonic point corresponds to the former/latter).
For radial perturbations, out of the two possible shock locations, the inner one is 
unstable due to post-shock acceleration, while the outer one is stable due to post-shock deceleration 
(\citealt{nakayama1992, nobuta1994})\footnote{While these papers refer to their perturbation analysis as axisymmetric, they only consider radial 
variations in the perturbations.}.
However, even the outer shock is unstable to non-axisymmetric perturbations. 
These instabilities
were invoked to explain time variability in accreting systems (e.g., \citealt{mtk1999}; see \citealt{Iwakami2009} for simulations of SASI with rotation). 
The occurrence of non-axisymmetric Papaloizou-Pringle instability and its interplay with the advective-acoustic cycle is an additional 
complication with rotation (\citealt{gu2003}). Moreover, MHD turbulence in the accretion flow (\citealt{Balbus1998}) may damp the advection 
of entropy and vorticity, an essential component of SASI.
\citet{Foglizzo2005} show that the advective-acoustic cycle also results in shock oscillations for wind accretion.
Although shock oscillations appear to be robust, for modeling observations they must be studied in a more realistic set up
than ours.

\section{Summary}
\label{sect:conclusions}
In this paper we study the dependence of spherical accretion on the nature of the central accretor. In particular, we study the influence of the 
hard surface of a neutron star (NS) which slows down the accreting matter to subsonic speeds. 
In contrast, spherically infalling matter onto a black hole (BH) accelerates to 
supersonic speeds and eventually is lost through the event horizon. Following are the key findings of our study:

\begin{itemize}
 \item We obtain similarity solutions for the flow profiles for accretion on to BHs (matter is allowed to accrete at supersonic speeds) and 
 NSs (matter comes to rest at the surface). While an outward-propagating shock is present for NSs, BH profiles are smooth and transonic.
 While similar work was done in the past, we have obtained the complete flow profiles, particularly the solutions outside the shock in case of
 NSs.
 
 \item Classical Bondi accretion theory is applicable only for adiabatic spherically symmetric inviscid accretion onto BHs, but not onto NSs. Due to the 
 presence of a surface, matter has to slow down at the surface, which gives rise to a shock. To study the effect of a surface, we study accretion
  with two different inner boundary conditions --  reflective and steady-shock. 
For the  reflective boundary condition, an outward propagating shock is launched, but ultimately there is no shock as it moves out and weakens. 
Entropy for the post-shock gas decreases with radius, and this subsonic flow profile is convectively unstable. Eventually an isentropic, hydrostatic,
hot atmosphere is formed in this case.
 For the steady-shock boundary condition, we allow a small subsonic infall velocity at the inner boundary. If the velocity at the inner boundary is
 within a certain range (see Fig. \ref{fig:vel_shock}), a steady standing shock is obtained in 1-D. However, in 2-D the standing shock is unstable
 to the standing accretion shock instability (SASI), giving rise to radial ($l=0$) and vertical ($l=1$) oscillations.
The effect of radial oscillations is reflected in the luminosity curve (Fig. \ref{fig:qpos}) which shows  coherent oscillations. This implies SASI can be 
a possible mechanism for quasi-periodic oscillations (QPOs) in XRBs (even in BHs in which a sudden deceleration of the infalling flow can be 
produced by a centrifugal barrier, rather than a surface).
 
 \item In our model, the velocity at the inner boundary (physically, this may be governed by cooling close to the inner surface) controls 
 different spherical accretion regimes. If cooling allows for the inflow velocity at the surface to be faster than a limit ($v_{\rm in} > v_{\rm in,min}$), accretion rate on to the NS, $\dot{M}_{NS} = \dot{M}_B$. But for $v_{\rm in} < v_{\rm in,min}$, the accretion rate,  $\dot{M}_{NS} < \dot{M}_B$. In both cases the amount of matter reaching the NS surface is less than that crossing the BH event horizon. This is because for Bondi accretion $\dot{M}_B \propto M^2$ ($M$ is the mass of the central accretor), and BHs are more massive than NSs.
 
 \item In the quiescent state, some NS XRBs are observed to be more luminous compared to BX XRBs. This has been interpreted as an evidence for
 the advection of the majority of energy released due to accretion across the event horizon of BHs. However, this argument assumes that the accretion
 rate on to BHs and NSs embedded in a similar environment are identical. This assumption is unlikely to be true in the quiescent state with inefficient
  cooling, as the accretion rate (scaled to the Bondi value) is likely to be much smaller for NSs, which have surfaces and an effective inflow velocity 
  close to the surface $v_{\rm in} < v_{\rm in, min}$ (see Fig. \ref{fig:vel_shock}). Spherical accretion is an appropriate model for the radiatively 
  inefficient quiescent accretion flow,
  which is optically thin but geometrically thick. A floor in X-ray luminosity of NS XRBs may be due to radiation not linked to the current accretion rate,
  but due to other effects such as thermal radiation from the NS surface. 
 
\end{itemize}

We aim to improve our -- admittedly simple -- models to include important physical effects such as angular momentum and magnetic fields. Our
paper provides a basic foundation for more realistic future simulations, which are needed to directly match with the observations of NS and BH XRBs. 

\section*{Acknowledgments}
We thank Ramesh Narayan, Dinshaw Balsara and Deepto Chakrabarty for helpful discussions. We also thank the referee Thierry Foglizzo for useful
suggestions. PD thanks Kartick Sarkar and Naveen Yadav for
technical help. PS acknowledges the Department of Science and Technology, India grant no. Sr/S2/HEP-048/2012 and an India-Israel
joint research grant (6-10/2014[IC]). BM acknowledges an Indo-Bulgarian Project funded by the
Department of Science and Technology, India, with
grant no. INT/BULGARIA/P-8/12.

\bibliographystyle{mn2e}
\bibliography{bibtex}
\label{lastpage}

\end{document}